\documentclass[fleqn,usenatbib]{mnras}

\usepackage{graphicx}	
\usepackage{amsmath}	
\usepackage{amssymb}	
\usepackage{ragged2e}
\usepackage{comment}
\usepackage{capt-of}
\usepackage{bigints} 
\usepackage{physics}
\usepackage{xcolor}
\usepackage{cleveref}
\usepackage{longtable}
\usepackage{xtab}
\usepackage{bm}
\usepackage{cancel}
\usepackage{CJK}
\usepackage{multirow}

\allowdisplaybreaks

\usepackage{scalerel,tikz}
\usetikzlibrary{svg.path}
\definecolor{orcidlogocol}{HTML}{A6CE39}
\tikzset{orcidlogo/.pic={
 \fill[orcidlogocol] svg{M256,128c0,70.7-57.3,128-128,128C57.3,256,0,198.7,0,128C0,57.3,57.3,0,128,0C198.7,0,256,57.3,256,128z};
 \fill[white] svg{M86.3,186.2H70.9V79.1h15.4v48.4V186.2z}
 svg{M108.9,79.1h41.6c39.6,0,57,28.3,57,53.6c0,27.5-21.5,53.6-56.8,53.6h-41.8V79.1z M124.3,172.4h24.5c34.9,0,42.9-26.5,42.9-39.7c0-21.5-13.7-39.7-43.7-39.7h-23.7V172.4z}
 svg{M88.7,56.8c0,5.5-4.5,10.1-10.1,10.1c-5.6,0-10.1-4.6-10.1-10.1c0-5.6,4.5-10.1,10.1-10.1C84.2,46.7,88.7,51.3,88.7,56.8z};
}}
\DeclareRobustCommand{\orcidicon}[1]{%
  \href{https://orcid.org/#1}{%
    \raisebox{-0.15ex}{%
      \resizebox{8pt}{8pt}{%
        \begin{tikzpicture}[yscale=-1,transform shape]
          \pic{orcidlogo};
        \end{tikzpicture}%
      }%
    }%
  }%
}

\defcitealias{Lancaster2026a}{L26a}
\defcitealias{Lancaster2026b}{L26b}
\defcitealias{Goldreich1995ApJ}{GS95}
\defcitealias{kolmogorov1941dissipation}{K41}

\title[Magnetized TRMLs]{How Magnetic Fields Regulate Cooling and Mixing in Turbulent Radiative Mixing Layers}

\author[Mohapatra et al.]{%
Rajsekhar Mohapatra\,\orcidicon{0000-0002-1600-7552}$^{1}$\thanks{E-mail: rmohapatra@princeton.edu (RM)},
Lachlan Lancaster\,\orcidicon{0000-0002-0041-4356}$^{2}$,
Drummond Fielding\,\orcidicon{0000-0003-3806-8548}$^{3}$,
\newauthor
Eliot Quataert\,\orcidicon{0000-0001-9185-5044}$^{1}$
and Greg L. Bryan\,\orcidicon{0000-0003-2630-9228}$^{4}$
\\
$^{1}$Department of Astrophysical Sciences, Princeton University, Princeton, NJ 08544, USA\\
$^{2}$Center for Computational Astrophysics, Flatiron Institute, 162 5th Avenue, New York, NY 10010, USA\\
$^{3}$Department of Physics, New York University, 726 Broadway, New York, NY, 10003, USA\\
$^{4}$Department of Astronomy, Columbia University, 550 W 120th Street, New York, NY 10027, USA
}
\date{Accepted XXX. Received YYY; in original form ZZZ}

\pubyear{\the\year{}}

\begin{document}
\label{firstpage}
\pagerange{\pageref{firstpage}--\pageref{lastpage}}
\maketitle

\begin{abstract}
Turbulent radiative mixing layers (TRMLs) are expected wherever hot and cold gas move past one another, including in the solar corona, galactic winds, and cold filaments in galaxy clusters. These environments are often magnetized, but magnetic effects on mixing and cooling remain less well understood than in the hydrodynamic (HD) case. We present magnetohydrodynamic (MHD) simulations of TRMLs at resolutions up to $1024 \times 2048^2$, spanning fields aligned with and transverse to the shear, and polarity-reversing configurations in which oppositely directed fields form current sheets at the interface. Even initially weak hot-phase fields, with $\mathcal{M}_{\rm A,shear}\equiv v_{\rm shear}/v_{\rm A}\sim14$, reduce the mass and enthalpy flux and radiative cooling rate by up to an order of magnitude relative to HD. Magnetic tension weakens turbulent motions, reducing both the diffusion of hot gas into the layer and the folding of the cooling surface. Transverse fields suppress cooling somewhat more strongly than shear-aligned ones, although a transverse field exerts no tension against the initial linear instability. Polarity reversal changes the morphology of the cooling gas without restoring HD-like mixing. The dependence on the Damk\"ohler number is similar to the HD case, but the degree of suppression is dependent on the initial field orientation. The net cooling rate appears resolution-independent in HD but declines with resolution in MHD, and has not converged, so our suppression factors are lower limits. Magnetic fields therefore strongly regulate cooling and mixing in multiphase gas, and quantitative predictions require careful treatment of transport processes.
\end{abstract}




\section{Introduction}\label{sec:introduction}


Multiphase gas, composed of coexisting components that differ by orders of magnitude in temperature and density, is ubiquitous in astrophysical plasmas. It is observed in environments ranging from the solar corona \citep{Antolin2020PPCF} and the multiphase interstellar medium \citep[ISM;][]{Cox2005ARAA} to the circumgalactic medium \citep[CGM;][]{Tumlinson2017review,Faucher-Giguere2023ARAA} and the intracluster medium \citep[ICM;][]{Werner2019SSRv}. This multiphase structure includes not only hot, volume-filling plasma and cold atomic and molecular filaments, but also gas at intermediate temperatures where radiative cooling is highly efficient. Observational tracers of this intermediate-temperature gas include coronal rain and transition-region emission in the solar atmosphere \citep[e.g.][]{Ishikawa2020SoPh,Sahin2023ApJ}, O\,{\sc vi}-bearing gas in the ISM, Galactic halo, and superbubbles \citep[e.g.][]{Savage2003ApJS,Barstow2010ApJ,Danforth2006ApJ,Sankrit2007PASP}, and O\,{\sc vi} absorption or emission in galactic winds \citep[e.g.][]{Grimes2009ApJS,Kim2024AJ}. Related intermediate-temperature gas is also observed in nearby galaxy haloes \citep{Tumlinson2013ApJ,HWChen2020MNRAS}, galaxy groups \citep{Olivares2022A&A}, and cluster cores \citep{Oegerle2001ApJ,Bregman2006ApJ}.

Turbulent radiative mixing layers (TRMLs) provide a natural mechanism for producing such gas. Inherited structure in the background velocity field along with a host of dynamical instabilities (Kelvin--Helmholtz [KH], Rayleigh--Taylor, Darrieus--Landau, etc.) can act to promote turbulence and mixing between the two phases. If the cooling time of the mixed gas is short, radiative losses allow this material to cool efficiently, allowing mixed gas to condense onto the cold phase. This process has been studied in the context of cold clouds in hot winds, turbulent multiphase media, and cold streams feeding galaxies, where cooling in the mixed gas can promote cold gas survival and growth, and help power observable line emission \citep{Gronke2018,Mohapatra2019,Sparre2020MNRAS,Mandelker2020MNRASa,Mohapatra2022MNRASb}. More generally, TRMLs have become an important framework for understanding mass, momentum, and energy exchange in multiphase environments \citep{Ji2019MNRAS,Fielding2020ApJ,Tan2021MNRASa,Tan2021MNRASb}.

Observational evidence also suggests that hot and cold gas can be closely coupled. In particular, the nearly linear relation between H$\alpha$ and X-ray emission in jellyfish galaxy tails and around cold filaments in galaxy clusters \citep{Sun2021,Olivares2025NatAs} indicates that gas at very different temperatures can be spatially and energetically linked. Such a correspondence can arise if emission from cold, intermediate-temperature, and hot gas is regulated by the same hot--cold boundary physics \citep{Chen2026arXivb}, with TRMLs continually replenishing the short-lived gas at intermediate temperatures.

While hydrodynamic (HD) TRMLs have received substantial attention, the role of magnetic fields remains less well understood. This is a significant omission because magnetic fields are observed across multiphase environments where hot and cold gas coexist. In the Milky Way, Faraday rotation and diffuse synchrotron emission reveal structured fields in the ionized ISM and halo, while Zeeman measurements show dynamically relevant fields in atomic and molecular gas \citep{Dickey2022ApJ,Seta2025MNRAS}. Faraday-rotation studies of high-velocity clouds infer line-of-sight fields of $\gtrsim 5$--$8\,\mu{\rm G}$ in the Smith high-velocity cloud \citep{Hill2013ApJ,Betti2019ApJ}, and fields of $\sim 6\,\mu{\rm G}$ and $\sim 0.3\,\mu{\rm G}$ in the Magellanic Leading Arm and Bridge, respectively \citep{McClure-Griffiths2010ApJ,Kaczmarek2017MNRAS,Jung2021MNRAS}. Beyond the Local Group, radio and submillimetre polarization observations of M82 reveal ordered fields associated with its galactic wind \citep{Adebahr2017A&A,Pattle2021MNRAS}, while \citet{Malik2026arXiv} report excess RM dispersion around galaxies hosting Mg\,{\sc ii} absorbers in the CGM. On cluster scales, Faraday rotation and diffuse synchrotron emission indicate ICM fields of order $\mu{\rm G}$, reaching $\sim 10$--$40\,\mu{\rm G}$ in some cool-core centres \citep{Carilli2002ARA&A,Govoni2004IJMPD,Feretti2012A&ARv}. In the Perseus cluster, large rotation measures toward 3C~84 and the narrow H$\alpha$ threads around NGC~1275 suggest magnetic support of cold filaments \citep{Taylor2006MNRAS,Fabian2008Nature}. These observations motivate studying TRMLs with magnetohydrodynamic (MHD) rather than purely HD models.

MHD simulations of hot--cold gas interactions have begun to clarify how magnetic fields modify multiphase flows across a range of scales. Cosmological MHD simulations and refined CGM zooms, including IllustrisTNG and GIBLE, show that magnetic fields are amplified by turbulence, outflows, and structure formation, and can shape multiphase gas on galactic and halo scales \citep{Marinacci2018MNRAS,Ramesh2024MNRAS,Ramesh2024A&A}. Complementary turbulent-box and CGM patch simulations show that cold gas survival depends sensitively on the competition between cooling and mixing, and that magnetic fields can alter the morphology and lifetime of the cold phase \citep{Das2024MNRAS,Mohapatra2022MNRASa,Mohapatra2025arXiv}. On smaller scales, cloud--wind simulations show that magnetic fields can drape around cold gas, suppress HD instabilities, and modify cloud survival, entrainment, and acceleration \citep{Dursi2008ApJ,Sparre2020MNRAS,Jung2023MNRASa,Cottle2020ApJ,Bruggen2023ApJ,HidalgoPineda2024MNRAS}.

Direct studies of magnetized TRMLs show that magnetic fields can be amplified within the mixing layer, suppress turbulent mixing, and alter the cooling and emission from intermediate-temperature gas \citep{Ji2019MNRAS,Das2024MNRAS,Zhao2023MNRAS}. Several key questions remain open. The first is how strongly the suppression depends on the orientation of the field relative to the shear, where existing results conflict. \citet{Ji2019MNRAS} found the suppression largely insensitive to the initial field direction, as did \citet{Zhao2023MNRAS} for fields lying in the plane of the hot--cold interface, though they found much stronger suppression when the field was initialized normal to it. \citet{Das2024MNRAS}, by contrast, found their transverse-field runs to be less strongly suppressed than their shear-aligned ones.

The second is that realistic multiphase environments may contain polarity reversals, so cold gas moving through a turbulent magnetized halo can encounter regions with oppositely directed fields, forming current sheets and reconnection-prone structures \citep{Fielding2023ApJ}. The third is numerical convergence, which bears directly on the first. Existing studies have used cell sizes of $\Delta x\simeq1/64$--$1/128$, with limited convergence testing at higher resolution. Part of the field amplification in the layer comes from small-scale dynamo action, which operates more efficiently as the numerical Reynolds number increases \citep[e.g.][]{Schekochihin2004ApJ,Federrath2011ApJ}. Both the saturated field strength near the interface and the geometry of the field that develops there may therefore still be changing with resolution, so the orientation dependence measured at low resolution need not be the one that survives.

The convergence issue is especially important because HD TRMLs can show apparently converged global quantities even when the small-scale cooling structure is unresolved. Recent work suggests that this apparent convergence arises from compensating resolution dependences: as the resolution increases, the hot--cold boundary area grows while the effective diffusion velocity across it decreases, yielding nearly unchanged global cooling rates despite unresolved microscopic transport \citep[][hereafter \citetalias{Lancaster2026a}]{Lancaster2026a}. Whether this compensation persists in MHD is not guaranteed. Magnetic fields introduce additional dynamically important scales, modify the geometry of hot--cold boundaries, and can change how the cooling surface area grows with resolution. These concerns have already been shown to be important in global wind-blown bubble calculations \citep{Lancaster2024ApJ}.

In this work, we carry out a systematic study of MHD TRMLs, focusing on how magnetic fields regulate mixing, cooling, and the structure of gas at intermediate temperatures. We explore different field orientations, polarity-reversing fields, and cases in which the hot and cold phases have different magnetizations. We also test whether the apparent convergence of global cooling and mixing rates found in HD carries over to MHD. Our simulations are designed to identify which magnetic configurations most strongly modify the structure and energetics of the mixing layer, and to determine when global cooling and mixing rates are numerically robust.

This paper is organized as follows. We describe the simulation setup and numerical methods in \S\ref{sec:Methods}. Results from the fiducial simulations are presented in \S\ref{sec:results-fid}, while the effects of varying the cooling rate and magnetic field strength are explored in \S\ref{sec:results-diff-xi-mag}. We present convergence tests in \S\ref{sec:diff_res}, discuss the implications of our results in \S\ref{sec:discussion}, and summarize our conclusions in \S\ref{sec:Conclusion}.


\section{Methods}\label{sec:Methods}


\subsection{Simulated Equations}\label{subsec:ModEq}

The evolution of the magnetized TRML is governed by the compressible ideal MHD equations,
\begin{subequations}
	\begin{align}
	\label{eq:continuity}
	&\frac{\partial\rho}{\partial t}+\nabla\cdot (\rho \mathbf{v})=0,\\
	\label{eq:momentum}
	&\frac{\partial(\rho\mathbf{v})}{\partial t}
	+\nabla\cdot \left(\rho \mathbf{v}\otimes \mathbf{v}+P^* I-\mathbf{B}\otimes \mathbf{B}\right)=0,\\
	\label{eq:energy}
	&\frac{\partial E}{\partial t}
	+\nabla\cdot \left[(E+P^*)\mathbf{v}-(\mathbf{B}\cdot\mathbf{v})\mathbf{B}\right]
	=Q-\mathcal{L},\\
	\label{eq:induction}
	&\frac{\partial\mathbf{B}}{\partial t}
	-\nabla\times(\mathbf{v}\times\mathbf{B})=0,\\
	\label{eq:pressure}
	&P^*=P+\frac{\mathbf{B}\cdot\mathbf{B}}{2},\\
	\label{eq:tot_energy}
	&E=\frac{1}{2}\rho\mathbf{v}\cdot\mathbf{v}
	+\frac{P}{\gamma-1}
	+\frac{1}{2}\mathbf{B}\cdot\mathbf{B}.
	\end{align}
\end{subequations}
Here, $\rho$ is the gas mass density, $\mathbf{v}$ is the velocity field, $\mathbf{B}$ is the magnetic field, and $P$ is the thermal pressure. The total pressure is denoted by $P^*$, and $E$ is the total energy density, including kinetic, thermal, and magnetic contributions. We assume an ideal-gas equation of state,
\begin{equation}
	P=\rho T,
\end{equation}
where $T$ is the gas temperature in units of velocity squared. The adiabatic index is set to $\gamma=5/3$. The source terms $Q$ and $\mathcal{L}$ denote volumetric heating and radiative cooling, respectively, which we describe below.

\subsection{Heating and cooling implementation}
\label{subsec:heating_cooling_implementation}
\begin{subequations}
We include heating and cooling as operator-split source terms in \cref{eq:energy}. The source terms maintain the cold phase and prevent the unperturbed hot phase from cooling, so that intermediate-temperature gas originates through mixing at the interface between the phases. We use an idealized prescription similar to \citet{ZChen2023ApJ}, also used in \citetalias{Lancaster2026a}. The cooling rate peaks at
\begin{equation}
	T_{\rm pk}=\left(T_{\rm cold}^2T_{\rm hot}\right)^{1/3},
\end{equation}
and is given by
\begin{equation}
	\label{eq:cooling_rate}
	\mathcal{L}(P,T)=\mathcal{L}_{\rm max}(P)
	\left(\frac{T}{T_{\rm pk}}\right)^{-\beta_{\rm cool}(T)},
\end{equation}
where
\begin{equation}
	\label{eq:cooling_rate_max}
	\mathcal{L}_{\rm max}(P)=
	\frac{1}{\gamma-1}\frac{\bar{P}}{t_{\rm cool,min}}
	\left(\frac{P}{\bar{P}}\right)^2 .
\end{equation}
Here, $\bar{P}$ is the initial hot-phase thermal pressure and $t_{\rm cool,min}$ is the minimum radiative cooling time at $P=\bar{P}$. At the cooling peak, \begin{equation}     t_{\rm cool}(P,T_{\rm pk})     \equiv \frac{P}{(\gamma-1)\mathcal{L}_{\rm max}(P)}     =t_{\rm cool,min}\frac{\bar{P}}{P}. \end{equation} We use
\begin{equation}
	\beta_{\rm cool}(T)=
	\begin{cases}
		-2, & T<T_{\rm pk},\\
		3, & T>T_{\rm pk}.
	\end{cases}
\end{equation}

The heating rate has the same pressure dependence and is written, over the temperature range relevant for our simulations, as
\begin{equation}
	\label{eq:heating_rate}
	Q(P,T)=C_{\rm h}\mathcal{L}_{\rm max}(P)
	\left(\frac{T}{T_{\rm pk}}\right)^{\alpha_{\rm h}},
\end{equation}
where
\begin{equation}
	\label{eq:alpha_heat}
	\alpha_{\rm h}=-\frac{4}{3},
	\qquad
	C_{\rm h}=\left(\frac{T_{\rm cold}}{T_{\rm pk}}\right)^{10/3}.
\end{equation}
These choices ensure that the underlying, uncut heating and cooling functions balance at both phase temperatures,
\begin{equation}
	Q(\bar{P},T_{\rm cold})=\mathcal{L}(\bar{P},T_{\rm cold}),
	\qquad
	Q(\bar{P},T_{\rm hot})=\mathcal{L}(\bar{P},T_{\rm hot}).
\end{equation}

In the simulations, both heating and cooling are disabled above
\begin{equation}
	T_{\rm cutoff}=0.85T_{\rm hot}.
\end{equation}
This cutoff prevents the unperturbed hot phase from cooling while allowing gas that has mixed and cooled away from the hot phase to radiate.
So the applied source term is
\begin{equation}
	\label{eq:applied_source_term}
	Q_{\rm app}-\mathcal{L}_{\rm app}
	=
	\begin{cases}
		Q(P,T)-\mathcal{L}(P,T), & T<T_{\rm cutoff},\\
		0, & T>T_{\rm cutoff}.
	\end{cases}
\end{equation}
\end{subequations}
The factor $(P/\bar{P})^2$ in equation~\eqref{eq:cooling_rate_max} mimics the density-squared scaling of collisional cooling. 

\subsection{Numerical methods}\label{subsec:numerical_methods}

We evolve the ideal MHD equations using \texttt{AthenaK}\footnote{\url{https://github.com/IAS-Astrophysics/athenak}}, a GPU-enabled, performance-portable version of \texttt{Athena++} \citep{Stone2020ApJS,Stone2026ApJS} built on the Kokkos library \citep{Trott2021CSE}. We use second-order RK2 time integration, piecewise-linear spatial reconstruction, the HLLD Riemann solver for MHD runs, and the HLLC solver for HD runs. We also apply a first-order flux-correction algorithm \citep{Lemaster2009ApJ} to handle cells with unphysical velocities or temperatures.

\subsection{Initial conditions}\label{subsec:initial_conditions}
\begin{subequations}
We simulate a cuboidal domain with dimensions
\begin{equation}
	(L_x,L_y,L_z)=(L_{\rm box},2L_{\rm box},2L_{\rm box}),
\end{equation}
using periodic boundary conditions in the $x$ and $y$ directions and outflow boundary conditions in the $z$ direction\footnote{\citetalias{Lancaster2026a} fix the top-z boundary values to the initial values, except for $v_z$. We find this configuration to be unstable for the MHD simulations.}. The extended domain in the $y$ direction, along the initial magnetic field in the shear-transverse (\texttt{By}) runs, accommodates longer-wavelength Alfv\'enic fluctuations. This choice is motivated by the anisotropy of magnetized turbulence, whose fluctuations can have larger correlation lengths along the magnetic field than across it \citep{Goldreich1995ApJ}. The larger extent in the $z$ direction keeps the hot--cold interface far from the outflow boundaries throughout the simulation. We set $L_{\rm box}=1$ in all runs.

Similar to \citetalias{Lancaster2026a}, we initialize the primitive variables using smooth tanh profiles across the interface at $z=0$. The hot phase has density and temperature
\begin{equation}
	\rho_{\rm hot}=1, \qquad T_{\rm hot}=1.
\end{equation}
We fix the temperature contrast to $100$ in all runs. Thus, in runs with thermal pressure equilibrium and no cold-phase magnetic-pressure enhancement, the cold phase has
\begin{equation}
	\rho_{\rm cold}=100, \qquad T_{\rm cold}=0.01.
\end{equation}

For simulations with uniform magnetic fields, we initialize the field in the desired direction with a specified plasma beta, $\beta$, so that
\begin{equation}
	B=\sqrt{\frac{2P}{\beta}} .
\end{equation}
For simulations with polarity-reversing fields, the magnetic field also transitions smoothly across the interface using a tanh profile. In runs where the magnetic field strength depends on density, including our fiducial polarity-reversing (PR) runs with fixed $B/\rho$, we keep the hot-phase magnetic field unchanged but increase the magnetic field strength in the cold gas. In these cases, we reduce the cold gas density, while keeping its temperature fixed, so that the hot and cold phases remain in total pressure equilibrium, including both thermal and magnetic pressure.

The asymptotic phase velocities are $\pm v_0$, chosen such that the Mach number is $0.25$ in the hot phase and $2.5$ in the cold phase. Thus,
\begin{equation}
    v_0=0.25\left(\sqrt{\frac{\gamma P}{\rho}}\right)_{\rm hot}.
\end{equation}
The velocity transitions smoothly across the mid-plane using a tanh profile. We define the shear velocity as the full velocity jump,
\begin{equation}
    v_{\rm shear}\equiv\Delta v=2v_0,
\end{equation}
and the shear time as
\begin{equation}
    t_{\rm shear}=\frac{L_x}{v_{\rm shear}}.
\end{equation}
We define the initial shear Alfv\'en Mach number as
\begin{equation}
    \label{eq:mach_alfven_shear}
    \mathcal{M}_{\rm A,shear}\equiv\frac{v_{\rm shear}}{v_{\rm A,hot}},
\end{equation}
where $v_{\rm A,hot}=B_{\rm hot}/\sqrt{\rho_{\rm hot}}$ is the initial hot-phase Alfv\'en speed.
We characterize the cooling strength by
\begin{equation}
    \xi\equiv\frac{t_{\rm shear}}{t_{\rm cool,min}}.
\end{equation}
Thus, larger values of $\xi$ correspond to more rapid cooling
relative to the shear time.

We seed the KH instability by adding weak turbulent velocity perturbations on small scales,
\begin{equation}
	8 \leq \frac{kL_{\rm box}}{2\pi} \leq 32 .
\end{equation}
These perturbations are several orders of magnitude smaller than the imposed shear velocity, so they seed the instability without significantly altering the initial shear flow.

\end{subequations}

\subsection{Simulation Suite}\label{subsec:list_of_simulations}

We perform a suite of simulations designed to explore how magnetic geometry, the mixing-to-cooling time ratio, and the strength of magnetization affect TRML evolution. The full list of runs is given in \Cref{tab:sim_params}. Our fiducial suite consists of five simulations with $\xi=15$ and $\beta_{\rm hot}=1000$. This suite includes one HD run, labelled \texttt{HD} ($\beta_{\rm hot} = \infty$); two uniform field MHD runs, labelled \texttt{Bx-U-q0} and \texttt{By-U-q0}; and two polarity reversing MHD runs, labelled \texttt{Bx-R-q1} and \texttt{By-R-q1}. In these labels, \texttt{Bx} and \texttt{By} denote the initial magnetic field direction, \texttt{U} denotes a uniform field, \texttt{R} denotes a polarity reversing field, and \texttt{q} gives the exponent in the imposed scaling $B\propto\rho^q$ across the phases.

The polarity reversing runs are motivated by the possibility that cold gas in a multiphase medium interacts with magnetic fields whose direction changes across the surrounding hot phase. Previous MHD TRML studies have shown that magnetic fields generally suppress turbulent mixing and reduce cooling by stabilizing the mixing layer \citep{Ji2019MNRAS,Zhao2023MNRAS,Das2024MNRAS}. However, a polarity reversal at the interface introduces magnetic reconnection at the hot--cold boundary. These runs therefore allow us to test whether reconnection and field rearrangement can offset the suppression of mixing by providing a dissipation mechanism for the field that is advected into the layer. This could alter the morphology of the cooling gas, or change the rate at which hot gas mixes with the cold phase.

We adopt the fixed-$B/\rho$ case as our fiducial polarity reversing configuration because it corresponds to magnetic flux conservation during one-dimensional compression. This choice is also motivated by multiphase MHD simulations in which cold gas can become strongly magnetized relative to the surrounding hot phase \citep[e.g.][]{Dursi2008ApJ,Jung2023MNRASa,Mohapatra2025arXiv}. The corresponding cold-to-hot density contrasts, determined from total pressure balance as described in \S\ref{subsec:initial_conditions}, are listed in \cref{tab:sim_params}. The highest resolution in the fiducial suite is $\Delta x=1/1024$ (we have dropped the $L_{\rm box}$ factor from here onwards).

We then vary the mixing-to-cooling time ratio by repeating the same five configurations with $\xi=1.5$, $3$, and $150$, at resolutions up to $\Delta x=1/1024$. For these runs, we append the value of $\xi$ to the fiducial labels, for example \texttt{HD-$\xi$1.5}, \texttt{Bx-U-q0-$\xi$1.5}, and \texttt{Bx-R-q1-$\xi$1.5}. We also test a more strongly magnetized hot phase with $\beta_{\rm hot}=100$, using the same magnetic configurations and appending \texttt{$\beta$100} to the labels.

Finally, we vary the cold phase magnetization in polarity reversing runs by considering fixed $B/\rho^{1/2}$ and fixed $B/\rho^{0.66}$, labelled \texttt{Bx-R-q1/2}, \texttt{By-R-q1/2}, \texttt{Bx-R-q2/3}, and \texttt{By-R-q2/3}, respectively. The \texttt{q2/3} labels denote the adopted exponent $q=0.66\simeq2/3$. The $q=1/2$ runs have equal Alfv\'en speed in the two phases, while the $q=0.66$ runs approximate the $B\propto\rho^{2/3}$ scaling expected from magnetic flux conservation during isotropic compression.

We list all of our simulations and some of their key parameters in \Cref{tab:sim_params}.
\begin{table*}
    \centering
    \def\arraystretch{1.25}
    \caption{Simulation parameters for the runs presented in this work. The fiducial suite compares HD, uniform-field MHD, and polarity-reversing MHD simulations at $\xi=15$ and $\beta_{\rm hot}=1000$. Additional suites vary the cooling time, magnetic field strength, and cold-phase magnetization.}
    \label{tab:sim_params}
    \resizebox{\textwidth}{!}{
        \begin{tabular}{llccccccc}
        \hline\hline
        Set
        & Label 
        & Resolution range 
        & $\xi$ 
        & $\beta_{\rm hot}$ 
        & $\rho_{\rm cold}/\rho_{\rm hot}$
        & Field direction 
        & $q$ in $B\propto\rho^q$ 
        & Polarity reversal \\
        {\scriptsize(1)}
        & {\scriptsize(2)} 
        & {\scriptsize(3)} 
        & {\scriptsize(4)} 
        & {\scriptsize(5)} 
        & {\scriptsize(6)}
        & {\scriptsize(7)} 
        & {\scriptsize(8)} 
        & {\scriptsize(9)} \\
        \hline\hline

        \multirow{5}{*}{Fiducial}
        & \texttt{HD} 
        & $1/128$--$1/1024$ 
        & $15$ 
        & $-$ 
        & $100$
        & $-$ 
        & $-$ 
        & No \\

        & \texttt{Bx-U-q0} 
        & $1/128$--$1/1024$ 
        & $15$ 
        & $1000$ 
        & $100$
        & $x$ 
        & $0$ 
        & No \\

        & \texttt{By-U-q0} 
        & $1/128$--$1/1024$ 
        & $15$ 
        & $1000$ 
        & $100$
        & $y$ 
        & $0$ 
        & No \\

        & \texttt{Bx-R-q1} 
        & $1/128$--$1/1024$ 
        & $15$ 
        & $1000$ 
        & $27.0$
        & $x$ 
        & $1$ 
        & Yes \\

        & \texttt{By-R-q1} 
        & $1/128$--$1/1024$ 
        & $15$ 
        & $1000$ 
        & $27.0$
        & $y$ 
        & $1$ 
        & Yes \\

        \hline

        \multirow{15}{*}{$\xi$ dependence}
        & \texttt{HD-$\xi$1.5} 
        & $1/128$--$1/1024$ 
        & $1.5$ 
        & $-$ 
        & $100$
        & $-$ 
        & $-$ 
        & No \\

        & \texttt{Bx-U-q0-$\xi$1.5} 
        & $1/128$--$1/512$ 
        & $1.5$ 
        & $1000$ 
        & $100$
        & $x$ 
        & $0$ 
        & No \\

        & \texttt{By-U-q0-$\xi$1.5} 
        & $1/128$--$1/1024$ 
        & $1.5$ 
        & $1000$ 
        & $100$
        & $y$ 
        & $0$ 
        & No \\

        & \texttt{Bx-R-q1-$\xi$1.5} 
        & $1/128$--$1/512$ 
        & $1.5$ 
        & $1000$ 
        & $27.0$
        & $x$ 
        & $1$ 
        & Yes \\

        & \texttt{By-R-q1-$\xi$1.5} 
        & $1/128$--$1/512$ 
        & $1.5$ 
        & $1000$ 
        & $27.0$
        & $y$ 
        & $1$ 
        & Yes \\

        & \texttt{HD-$\xi$3} 
        & $1/128$--$1/512$ 
        & $3$ 
        & $-$ 
        & $100$
        & $-$ 
        & $-$ 
        & No \\

        & \texttt{Bx-U-q0-$\xi$3} 
        & $1/128$--$1/512$ 
        & $3$ 
        & $1000$ 
        & $100$
        & $x$ 
        & $0$ 
        & No \\

        & \texttt{By-U-q0-$\xi$3} 
        & $1/128$--$1/512$ 
        & $3$ 
        & $1000$ 
        & $100$
        & $y$ 
        & $0$ 
        & No \\

        & \texttt{Bx-R-q1-$\xi$3} 
        & $1/128$--$1/512$ 
        & $3$ 
        & $1000$ 
        & $27.0$
        & $x$ 
        & $1$ 
        & Yes \\

        & \texttt{By-R-q1-$\xi$3} 
        & $1/128$--$1/512$ 
        & $3$ 
        & $1000$ 
        & $27.0$
        & $y$ 
        & $1$ 
        & Yes \\

        & \texttt{HD-$\xi$150} 
        & $1/128$--$1/1024$ 
        & $150$ 
        & $-$ 
        & $100$
        & $-$ 
        & $-$ 
        & No \\

        & \texttt{Bx-U-q0-$\xi$150} 
        & $1/128$--$1/512$ 
        & $150$ 
        & $1000$ 
        & $100$
        & $x$ 
        & $0$ 
        & No \\

        & \texttt{By-U-q0-$\xi$150} 
        & $1/128$--$1/512$ 
        & $150$ 
        & $1000$ 
        & $100$
        & $y$ 
        & $0$ 
        & No \\

        & \texttt{Bx-R-q1-$\xi$150} 
        & $1/128$--$1/512$ 
        & $150$ 
        & $1000$ 
        & $27.0$
        & $x$ 
        & $1$ 
        & Yes \\

        & \texttt{By-R-q1-$\xi$150} 
        & $1/128$--$1/1024$ 
        & $150$ 
        & $1000$ 
        & $27.0$
        & $y$ 
        & $1$ 
        & Yes \\

        \hline

        \multirow{4}{*}{Stronger field}
        & \texttt{Bx-U-q0-$\beta$100} 
        & $1/128$--$1/512$ 
        & $15$ 
        & $100$ 
        & $100$
        & $x$ 
        & $0$ 
        & No \\

        & \texttt{By-U-q0-$\beta$100} 
        & $1/128$--$1/512$ 
        & $15$ 
        & $100$ 
        & $100$
        & $y$ 
        & $0$ 
        & No \\

        & \texttt{Bx-R-q1-$\beta$100} 
        & $1/128$--$1/512$ 
        & $15$ 
        & $100$ 
        & $9.6$
        & $x$ 
        & $1$ 
        & Yes \\

        & \texttt{By-R-q1-$\beta$100} 
        & $1/128$--$1/512$ 
        & $15$ 
        & $100$ 
        & $9.6$
        & $y$ 
        & $1$ 
        & Yes \\

        \hline

        \multirow{4}{*}{cold gas magnetization}
        & \texttt{Bx-R-q1/2} 
        & $1/128$--$1/512$ 
        & $15$ 
        & $1000$ 
        & $91.0$
        & $x$ 
        & $1/2$ 
        & Yes \\

        & \texttt{By-R-q1/2} 
        & $1/128$--$1/512$ 
        & $15$ 
        & $1000$ 
        & $91.0$
        & $y$ 
        & $1/2$ 
        & Yes \\

        & \texttt{Bx-R-q2/3} 
        & $1/128$--$1/512$ 
        & $15$ 
        & $1000$ 
        & $71.9$
        & $x$ 
        & $0.66$ 
        & Yes \\

        & \texttt{By-R-q2/3} 
        & $1/128$--$1/512$ 
        & $15$ 
        & $1000$ 
        & $71.9$
        & $y$ 
        & $0.66$ 
        & Yes \\

        \hline\hline
        \end{tabular}
    }
    \vspace{0.8em}
    \justifying{\footnotesize 
    Notes: 
    
    We group the simulations by the parameter varied relative to the fiducial suite in Column (1), and give the labels used in figures and text in Column (2). The labels encode the magnetic field configuration: \texttt{HD} denotes hydrodynamics, \texttt{Bx} and \texttt{By} give the initial field direction, \texttt{U} denotes a uniform field, \texttt{R} denotes a polarity-reversing field, and \texttt{q} gives the exponent in $B/\rho^q$. We omit the fiducial values $\xi=15$ and $\beta_{\rm hot}=1000$ from the labels, and append non-fiducial values using \texttt{$\xi$} and \texttt{$\beta$}. We give the resolution range in cell size in Column (3). We give the mixing-to-cooling time ratio $\xi$ in Column (4), and the hot-phase plasma beta in Column (5). Column (6) gives the cold-to-hot density contrast, $\rho_{\rm cold}/\rho_{\rm hot}$. We set the temperature contrast to $T_{\rm hot}/T_{\rm cold}=100$ in all simulations, and choose $\rho_{\rm cold}/\rho_{\rm hot}$ so that the hot and cold phases are initially in total pressure equilibrium. Columns (7) and (8) give the field direction and magnetization scaling, respectively. Column (9) indicates whether the field reverses polarity across the hot--cold interface. All simulations use $L_{\rm box}=1$.
    }
\end{table*}

\section{Results: Fiducial runs}\label{sec:results-fid}

\subsection{2D slices}\label{subsec:2d_slice_fid}

\begin{figure*}
    \centering
    \includegraphics[width=\textwidth]{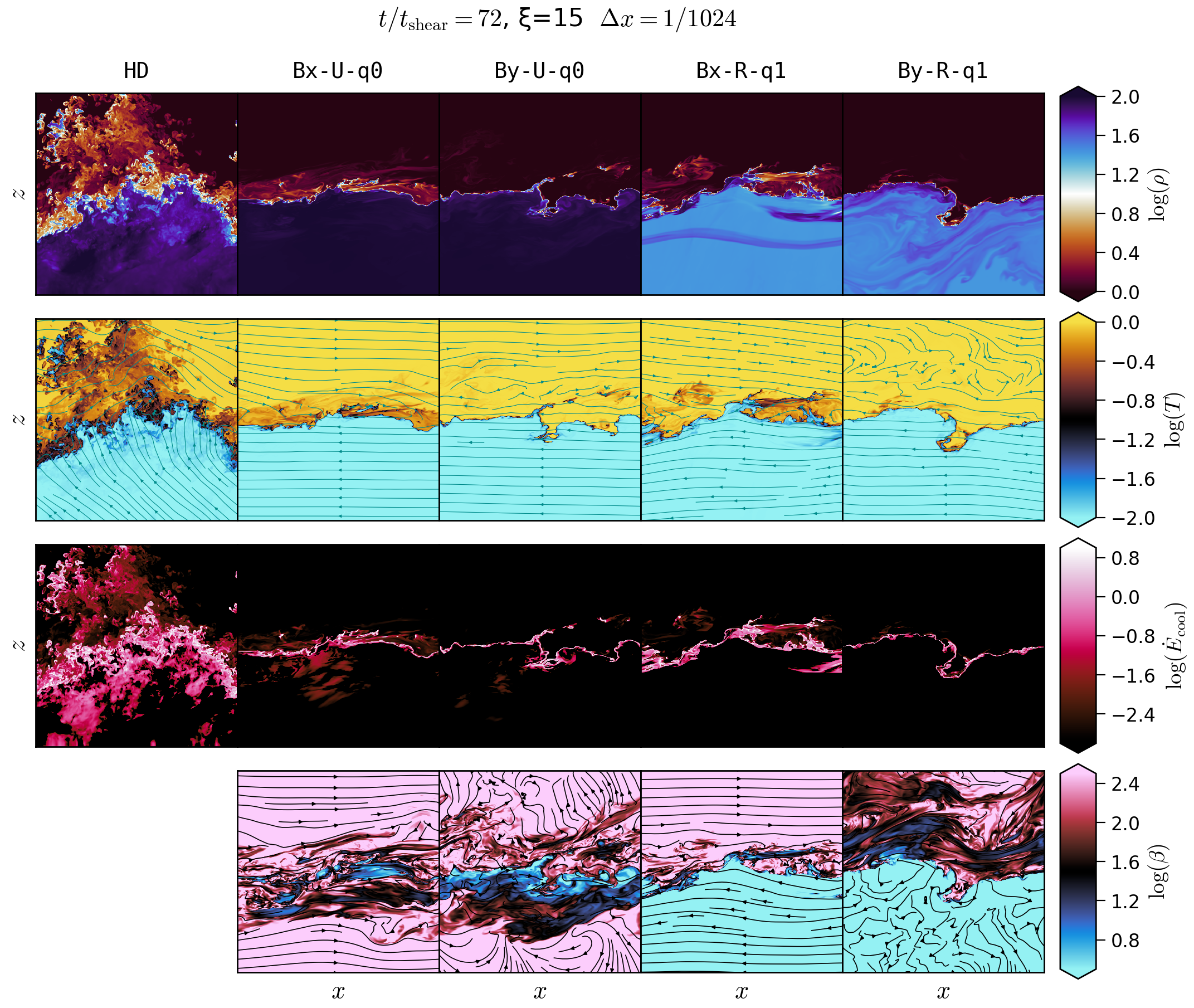}
    \caption{Slices in the $xz$-plane for the fiducial simulations at $t=72\,t_{\rm shear}$. From top to bottom, we show the logarithms of gas density, temperature, cooling rate, and plasma beta. The streamlines in the temperature row show the in-plane velocity field, while those in the plasma-beta row show the in-plane magnetic field. The \texttt{HD} run develops a broad turbulent layer with extended intermediate-temperature gas and strong hot gas inflow. In contrast, all MHD runs confine cooling to a thinner layer, show weaker inflow, and amplify the magnetic field near the interface. The fiducial runs have $\xi=15$ and $\beta_{\rm hot}=1000$. An animated version of this figure is available in the online supplementary material. The intermittent magnetic reconnection structures, which are difficult to observe in the $xz$-plane, are more readily apparent in \Cref{fig:vol_rendering_fid}.}
    \label{fig:slices_fid}
\end{figure*}

\Cref{fig:slices_fid} shows slices through the fiducial simulations at $y=0$. Although the domain extends over $-1<z<1$, we show only the region $-0.5<z<0.5$ around the hot--cold interface. The five columns correspond to the HD run, \texttt{HD}; the uniform-field runs, \texttt{Bx-U-q0} and \texttt{By-U-q0}; and the polarity-reversing runs, \texttt{Bx-R-q1} and \texttt{By-R-q1}. The first three rows show density, temperature, and cooling rate, while the fourth row shows the plasma beta for the MHD runs.

The HD run develops a broad, turbulent mixing layer with substantial gas at intermediate densities and temperatures. The cooling is distributed over a vertically extended region around the interface, rather than being confined to a narrow front. The velocity streamlines show a clear inflow of hot gas toward the mixing layer, consistent with the hot-phase enthalpy flux balancing the energy lost through radiative cooling at the mixing layer.

All the MHD runs differ qualitatively from the HD case. In all magnetic configurations, the intermediate-temperature gas is confined to a much thinner layer, and the cooling is concentrated near this thin hot--cold boundary, similar to the findings of \cite{Ji2019MNRAS,Zhao2023MNRAS,Das2024MNRAS}. The velocity streamlines show much weaker inflow toward the interface than in the HD run, consistent with reduced mixing and lower radiative losses. The polarity-reversing runs, \texttt{Bx-R-q1} and \texttt{By-R-q1}, also have a lower cold gas density than the uniform-field runs because the cold phase has a larger magnetic-pressure support in these configurations. Thus, their density contrast is smaller than the temperature contrast, although all fiducial runs have $T_{\rm hot}/T_{\rm cold}=100$.

The plasma-beta slices show that all MHD runs amplify the magnetic field near the interface, producing a localized decrease in $\beta$. This amplification is driven by compression as intermediate-temperature gas cools and by shear-induced motions at the interface, which stretch and fold the magnetic field until magnetic tension becomes strong enough to suppress further growth of the KH instability. Away from the interface, the \texttt{Bx} runs retain comparatively smooth field lines, indicating that the shear instabilities and the associated magnetic field amplification remain localized to the mixing layer. This localized field amplification also happens in the \texttt{By} runs, although this effect is more clearly visible in \Cref{fig:vol_rendering_fid,fig:vert-fid}.

\subsection{Volume rendering of fast-cooling gas}
\label{subsec:vol_rendering_fid}

\begin{figure*}
    \centering
    \includegraphics[width=0.75\textwidth]{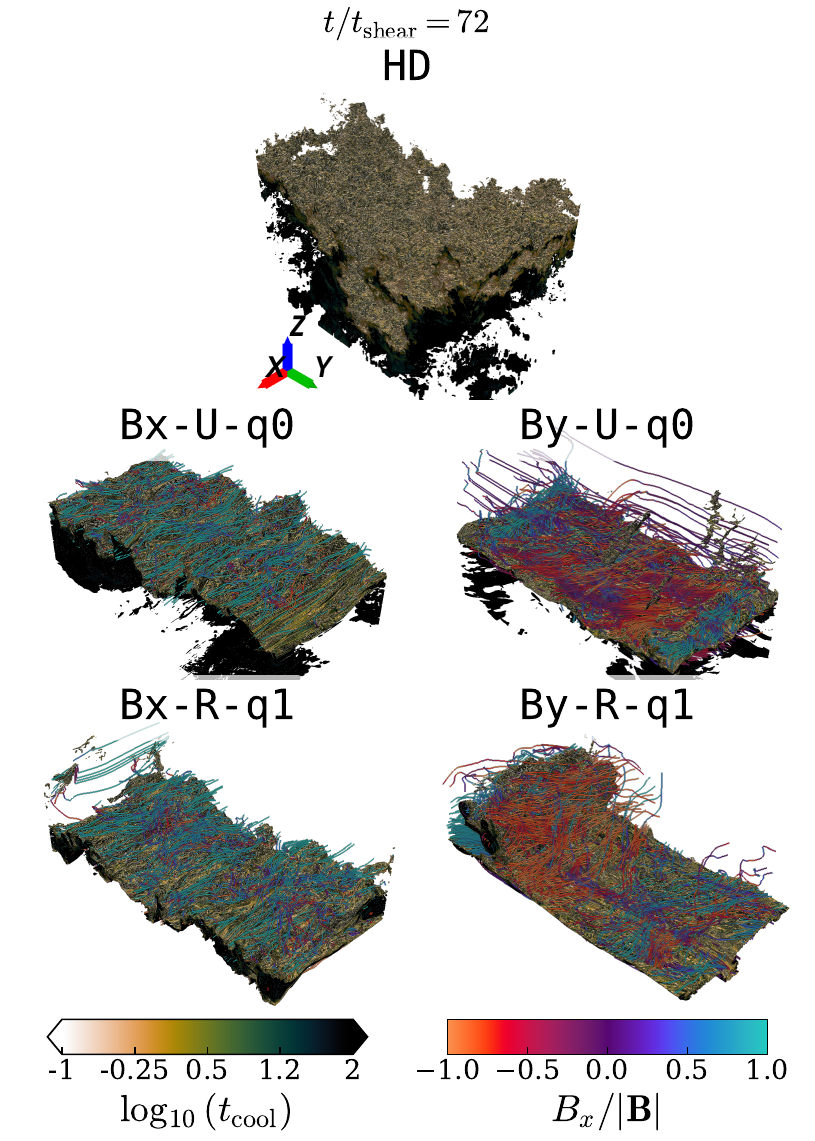}
    \caption{Volume rendering of gas at $T\leq T_{\rm pk}$ at $t=72\,t_{\rm shear}$ for the fiducial simulations with $\xi=15$ and $\beta_{\rm hot}=1000$. Gas with $T>T_{\rm pk}$ is transparent, while gas with $T\leq T_{\rm pk}$ is opaque and colored by the logarithm of the local cooling time. For the MHD runs, streamlines show magnetic field lines passing through gas near $T_{\rm pk}$, colored by their alignment with the $x$ direction. The \texttt{HD} run forms a highly wrinkled cooling surface, while the MHD runs have smoother and more ordered cooling gas. The \texttt{By} runs show the strongest suppression of small scale structure, and the polarity-reversing case \texttt{By-R-q1} produces a flux-tube-like morphology at the interface.}
    \label{fig:vol_rendering_fid}
\end{figure*}

In \Cref{fig:vol_rendering_fid}, we show volume renderings of gas at or below the cooling peak, $T\leq T_{\rm pk}$, with hotter gas made transparent. The color indicates the local cooling time, $t_{\rm cool}$. For the MHD runs, we also show magnetic field lines passing through gas near the hot--cold boundary, colored by $B_x/|\mathbf{B}|$.

Similar to \Cref{fig:slices_fid}, the \texttt{HD} run develops a highly folded cooling surface. Turbulence driven by the KH instability greatly enhances the area of gas near $T_{\rm pk}$ relative to the initial planar area, $L_xL_y$ \citep{Fielding2020ApJ}. The surface is deformed over a wide range of scales, from box-scale down to the grid scale. This morphology is consistent with the picture that, in an HD TRML with fast cooling ($\xi=15$ here), most cooling occurs along a corrugated hot--cold boundary whose effective area grows with resolution \citepalias{Lancaster2026a}.

The MHD runs show a much smoother cooling surface. The folds are fewer, broader, and mostly confined to large scales, indicating that magnetic tension suppresses the turnover of small-scale eddies. In the \texttt{Bx} runs, the magnetic field remains coherent away from the mixing layer, while gas near $T_{\rm pk}$ forms elongated structures that preferentially align with the mean field direction. This suggests that hot gas enters the cooling layer more anisotropically than in the HD case. The morphology is dominated by broad interface distortions rather than a cascade of small grid-scale folds.

The \texttt{By} runs show even less enhancement of the cooling-surface area. Although the initial field is along the $y$ direction, shear near the interface reorients part of the field toward the positive and negative $x$ directions. The presence of dynamically significant field components along both $x$ and $y$ further suppresses mixing at the interface. This reorientation also produces adjacent regions of oppositely directed field associated with current sheets. The effect is clearest in the polarity-reversing run, \texttt{By-R-q1}, where gas near $T_{\rm pk}$ is organized into elongated flux-tube-like structures along the shear direction. In an animated version of this visualization (available in the online version of the manuscript), these structures appear to originate from small-scale reconnecting regions and then merge into larger coherent features, reducing magnetic curvature and tension.

\subsection{Time evolution}
\label{subsec:time_evol_fid}

\begin{figure}
    \centering
    \includegraphics[width=0.5\textwidth]{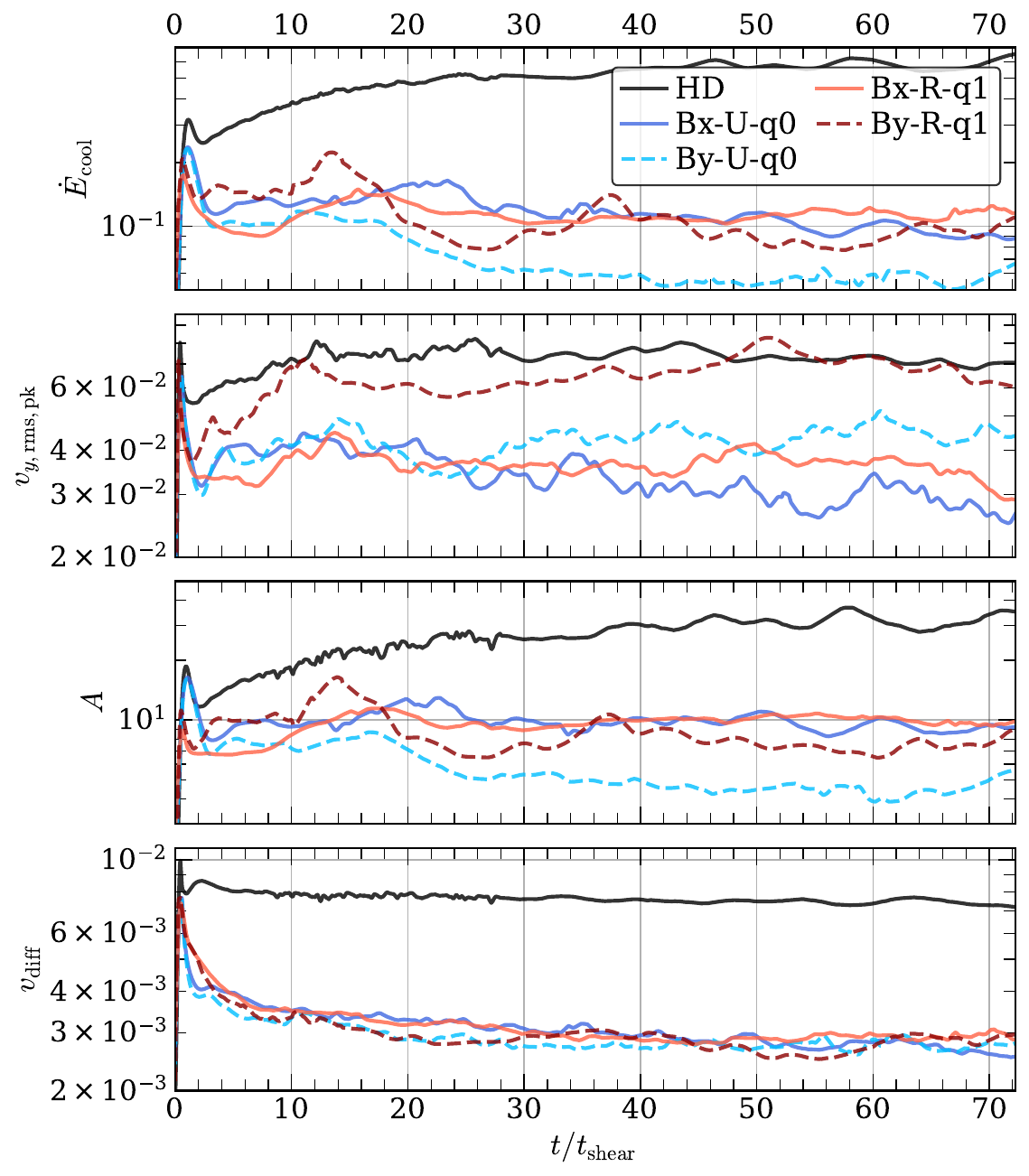}
    \caption{Time evolution of the net cooling rate, rms $v_y$ near the cooling peak, area of the $T_{\rm pk}$ surface, and diffusion velocity across it for the fiducial runs. In the saturated state, magnetic fields reduce the cooling rate by a factor of $\sim5$--$10$, the area by $\sim2$--$4$ and $v_{\rm diff}$ by $\sim3$. Among the MHD runs, the rms $v_y$ is largest in the \texttt{By} runs, where these motions are field-aligned.}
    \label{fig:time-evolution-fid}
\end{figure}

We show the time evolution of the net cooling rate, the rms $v_y$ near the cooling peak, the surface area of gas at $T_{\rm pk}$, $A$, and the diffusion velocity across the hot--cold interface, $v_{\rm diff}$, in \Cref{fig:time-evolution-fid}. The net cooling rate provides a measure of the total emission from rapidly cooling gas. The surface area and diffusion velocity are useful diagnostics because their product measures the effective volume flux across the interface near the cooling peak. When multiplied by the appropriate specific energy or density, this quantity gives estimates of the enthalpy flux and mass flux, which determine whether the cold phase gains or loses thermal energy and mass \citep{Fielding2020ApJ,Tan2021MNRASa}.

In a statistically steady TRML, we expect radiative losses at the interface to be balanced primarily by an enthalpy flux from the hot phase, provided that dissipative heating and heat-transfer processes, such as viscous, conductive, or resistive heating, are subdominant. In order to calculate $A$ and $v_{\rm diff}$ we follow the measurement procedure described in \S4 of \citetalias{Lancaster2026a} and summarize it here. We compute the surface area using the marching-cubes algorithm of \citet{Lewiner03_marching_cubes}, taking $T=T_{\rm pk}$ as the iso-value that identifies the peak-cooling surface. For the diffusion velocity, we define the direction into the cooling layer as 
\begin{subequations}
\begin{equation}
\label{eq:isosurface_normal}
\hat{\mathbf{n}}=-\nabla T/|\nabla T|, 
\end{equation}    
evaluated at $T=T_{\rm pk}$, and the local velocity into the layer as 
\begin{equation}
\label{eq:v_in}
v_{\rm in}=\mathbf{v}\cdot\hat{\mathbf{n}}.
\end{equation} 
The diffusion velocity is then the change in $v_{\rm in}$ across the layer,
\begin{equation}
    \label{eq:vdiff_measure}
    v_{\rm diff} = -\lambda\,\hat{\mathbf{n}}\cdot\nabla v_{\rm in},
\end{equation}
\end{subequations}
where $\lambda$ is a length representative of the thickness of the layer. Following \citetalias{Lancaster2026a} we evaluate this for all gas within a factor $10^{0.1}$ of $T_{\rm pk}$ and take $\lambda=1.2\,\Delta x$, so that both $A$ and $v_{\rm diff}$ are measured on the grid scale.\footnote{Our simulations include no explicit conduction, viscosity, or resistivity, so the hot and cold phases are mixed by numerical diffusion, which operates at the grid scale. In the absence of a resolved conduction scale, $v_{\rm diff}$ therefore measures the rate at which turbulence delivers gas to the smallest resolved scale, rather than a physical diffusion speed. This is why $v_{\rm diff}$ correlates with the large-scale turbulent velocity, as shown in \citetalias{Lancaster2026a} and in \S\ref{sec:appendix_enthalpy}, and also why it is expected to depend on resolution, an issue we return to in \S\ref{sec:diff_res}.}

During the first $\sim2\,t_{\rm shear}$, all runs, including the MHD ones, have cooling rates, rms $v_y$, interface areas, and diffusion velocities comparable to \texttt{HD}. The hot-phase fields are initially weak, with $\mathcal{M}_{\rm A,shear}\simeq14$, so the layer begins essentially hydrodynamic in character and the Kelvin--Helmholtz instability develops much as it does in \texttt{HD}. Compression in the cooling gas and shear-driven stretching then amplify the field near the interface. Once magnetic tension becomes dynamically important it resists the motions that fold the cooling surface: $v_{\rm diff}$ in the MHD runs falls below the \texttt{HD} value within a few shear times, and their interface areas stop growing while the \texttt{HD} area keeps increasing for $\sim30\,t_{\rm shear}$.

This transient has two consequences for how magnetized mixing layers should be simulated and compared. First, the approach to a magnetically saturated state is slow: the cooling rate of \texttt{By-U-q0} continues to decline secularly for up to $40\,t_{\rm shear}$. A simulation stopped after a few shear times would find the magnetic suppression to be much weaker than its asymptotic value. Second, part of the amplification responsible for this decline comes from small-scale dynamo action, which is known to operate more efficiently at higher numerical resolution \citep[e.g.][]{Schekochihin2004ApJ,Federrath2011ApJ}. At lower resolution the field grows more slowly, so the transient lasts longer and the saturated suppression is weaker (\Cref{fig:cool_rate_uni_diff_res}). Run time and resolution therefore both need to be taken into account when comparing suppression factors between studies, and we return to the resolution dependence in \S\ref{sec:diff_res}.

Once the runs reach a statistically steady state ($t/t_{\rm shear}=40$ onwards), all three of the cooling diagnostics are suppressed in the MHD runs relative to \texttt{HD}. The net cooling rate is reduced by a factor of $\sim5$--$10$, the area of the $T=T_{\rm pk}$ iso-surface by a factor of $\sim2$--$4$, and the diffusion velocity by a factor of $\sim2$--$3$. Given the interpretation of $v_{\rm diff}$ above, this indicates that magnetic fields strongly suppress the turbulent transport that feeds the cooling layer, consistent with previous MHD TRML studies \citep{Ji2019MNRAS,Das2024MNRAS,Zhao2023MNRAS}. Weaker turbulence also reduces the corrugation of the $T_{\rm pk}$ surface, leading to a smaller cooling area and hence a lower net cooling rate.

Interpreting the trends in rms $v_y$ requires more care, and they do not follow the same ordering. In HD, $v_y$ can be seen as a natural tracer of the turbulent velocity, since it is set neither by the imposed shear nor by the bulk inflow \citepalias{Lancaster2026a}. In MHD, no velocity component provides an unbiased measure of the turbulence, because the magnetic field breaks the symmetry between the remaining directions. In the \texttt{By} runs, $y$ is the initial field direction, so motions in $y$ are field-aligned and are not resisted by magnetic tension, and $v_{y,\rm rms}$ is correspondingly large. In the \texttt{Bx} runs, $y$ is perpendicular to the mean field, the same motions are opposed by tension, and $v_{y,\rm rms}$ is smaller. The effect is large enough that \texttt{By-R-q1} has an rms $v_y$ close to the \texttt{HD} value while cooling several times more slowly. The ordering of $v_{y,\rm rms}$ among the MHD runs therefore reflects the orientation of the field relative to $y$ at least as much as the strength of the turbulence, and should not be read directly as a measure of mixing. For this reason we do not use $v_{y,\rm rms}$ alone as the turbulent velocity when defining the Damk\"ohler number ($\mathrm{Da}$ is the ratio of the eddy turnover time to the minimum cooling time) in \S\ref{subsec:diff_xi}, but combine the two non-shear components after subtracting their mean profiles, as described in \S\ref{subsec:sigma_components_fid}.

Among the MHD runs, \texttt{By-U-q0} shows the strongest suppression of the cooling rate and interface area, despite retaining a relatively large $v_{y,\rm rms}$. The polarity-reversing runs, \texttt{Bx-R-q1} and \texttt{By-R-q1}, have cooling rates, interface areas, and diffusion velocities similar to \texttt{Bx-U-q0}, despite their distinct morphology.

In the following subsections, we present several diagnostics of the fiducial simulations: vertical profiles along the temperature-gradient direction in \S\ref{subsec:vert_profiles_fid}, probability distribution functions measured near the interface in \S\ref{subsec:pdfs_interface_fid}, component-wise velocity and magnetic field dispersions in \S\ref{subsec:sigma_components_fid}, and the fractal dimension of the interface in \S\ref{subsec:fractal_structure_fid}. Unless stated otherwise, these diagnostics, as well as the subsequent analyses of their dependence on different rates of cooling, different magnetic field strengths, and resolution, are computed in the statistically steady state, time-averaged over $t\geq 40\,t_{\rm shear}$.

\subsection{Vertical profiles}
\label{subsec:vert_profiles_fid}

\begin{figure*}
    \centering
    \includegraphics[width=\textwidth]{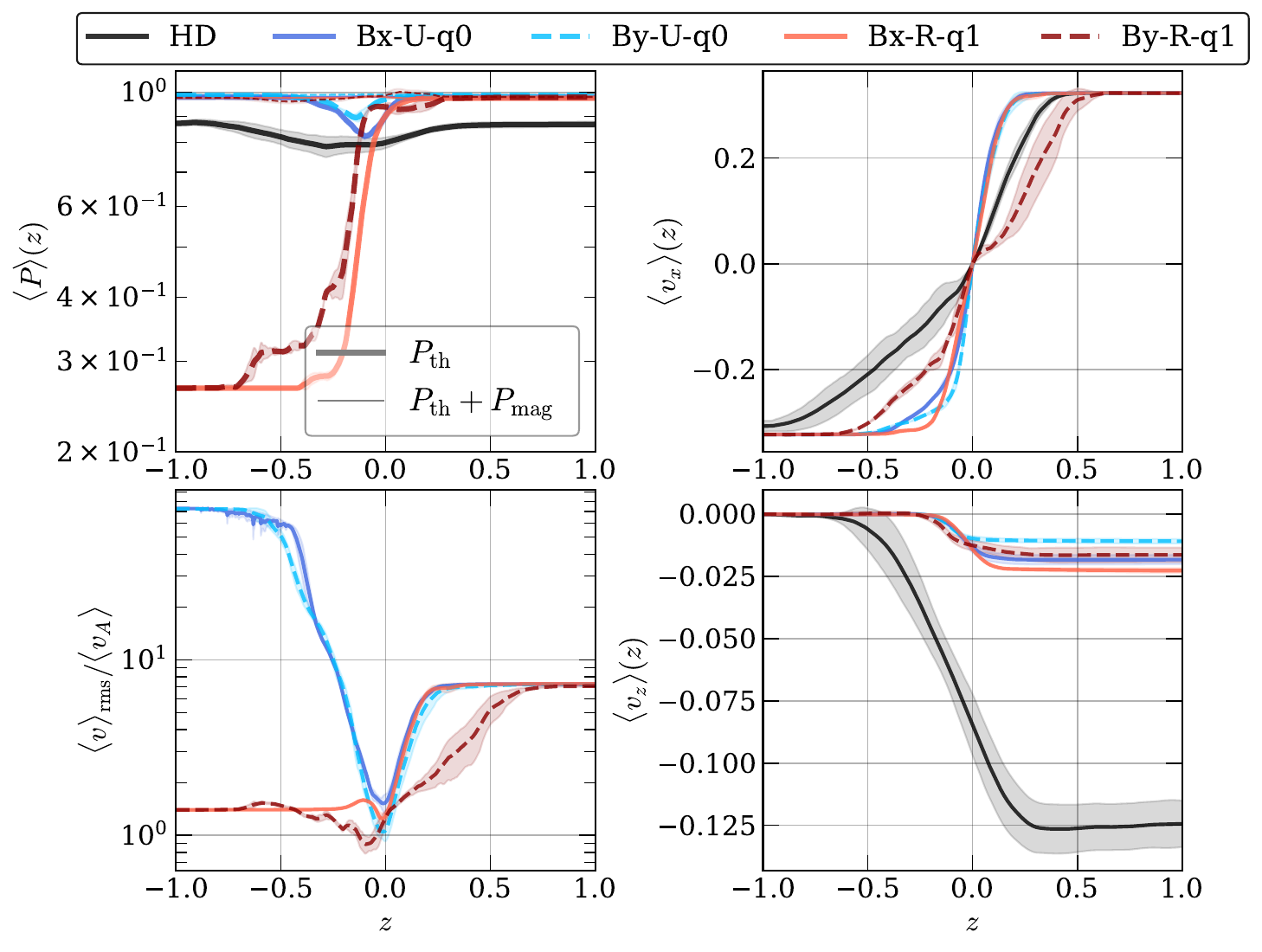}
    \caption{Vertical profiles for the fiducial simulations, averaged over the $xy$ plane and over the statistically steady state. The left column shows the thermal pressure (thinner lines add the magnetic pressure) and the Alfv\'en Mach number. The right column shows the velocity components along the shear, $v_x$, and along the temperature gradient, $v_z$. Shaded regions show the temporal variation. The MHD runs have sharper shear layers and much weaker hot gas inflow than \texttt{HD}. Magnetic pressure compensates the thermal-pressure dip near the interface, and the Alfv\'en Mach number decreases to unity near the layer, indicating dynamically important magnetic fields.}
    \label{fig:vert-fid}
\end{figure*}

In \Cref{fig:vert-fid}, we show $xy$-averaged vertical profiles for the fiducial simulations, further averaged over snapshots in the statistically steady state. Before averaging, each snapshot is shifted vertically to place the zero crossing of the mean $v_x$ profile nearest the domain centre at $z=0$. We also subtract the mean $v_z$ at the original height $z\simeq-L_{\rm box}$. The left column shows the thermal and total (thermal plus magnetic) pressure, and the Alfv\'en Mach number $\langle v\rangle_{\rm rms}/\langle v_A\rangle$. The right column shows the velocity components $v_x$ and $v_z$, along the shear and the temperature gradient, respectively. The shaded regions indicate the $1\sigma$ temporal variation in the statistically steady state.

The pressure profiles provide a useful diagnostic of the momentum balance across the layer. In the \texttt{HD} run, the thermal pressure shows a small dip near the interface, consistent with the quasi-steady TRML picture in which rapid cooling produces a thermal-pressure deficit that is partly balanced by the vertical Reynolds normal stress, $R_{zz}$ \citep{Sharma2025arXiv}. A related discussion of the pressure balance and Reynolds-stress decomposition is given in Appendix C of \citetalias{Lancaster2026a}. In the MHD runs, turbulence is substantially weaker, so the Reynolds-stress contribution is expected to be smaller. Instead, the thermal-pressure dip is compensated by magnetic pressure: the thin lines in the top-left panel show the sum of thermal and magnetic pressure, which remains nearly flat as a function of $z$.

The same compensation sets the difference in cold-phase pressure between the runs. Because the initial conditions enforce total pressure balance between the two phases, cold gas that carries part of its support magnetically must sit at lower thermal pressure, and hence lower density, at fixed temperature contrast. The polarity-reversing runs, \texttt{Bx-R-q1} and \texttt{By-R-q1}, have the most strongly magnetized cold phase and therefore the lowest cold-phase thermal pressure. We return to this in \S\ref{subsec:pdfs_interface_fid}, where the distributions show that it produces two distinct populations of cold gas.

The Alfv\'en Mach number profiles show that the magnetic field is amplified near the mixing layer in all MHD runs, consistent with previous MHD TRML studies \citep{Ji2019MNRAS,Zhao2023MNRAS,Das2024MNRAS}. The associated decrease in $\mathcal{M}_{\rm A}$ indicates that magnetic forces become dynamically important close to the interface. In the uniform-field runs, the mean Alfv\'en Mach number drops to $\sim1$ near the mixing layer. Cooling-induced compression and shear-driven stretching both contribute to this amplification. In the polarity-reversing runs, the mean $\mathcal{M}_{\rm A}\lesssim2$ throughout the cold phase, suggesting that the strongly magnetized cold gas resists shear-driven folding and mixing.

The velocity profiles establish the two geometric facts that the rest of this section relies on. First, the MHD shear layers are much narrower: $v_x$ transitions between the two phases over a smaller range in $z$ than in \texttt{HD}\footnote{The \texttt{By-R-q1} case is a small exception to this, likely due to the larger scale flux-tube like reconnection structures that develop.}, mirroring the sharper density and temperature transitions visible in the slices of \Cref{fig:slices_fid}. Second, the hot gas inflow is strongly reduced. Although $v_z$ is negative near the positive-$z$ side in all simulations, its magnitude is almost an order of magnitude larger in \texttt{HD} than in the MHD runs. Weaker turbulence reduces both the surface area of fast-cooling gas and the diffusion velocity across it, lowering the enthalpy supply to the cooling layer. The weaker bulk inflow and lower cooling rate are consistent with this reduced transport in steady state. We quantify the fluctuating motions behind this inflow in \S\ref{subsec:sigma_components_fid}.

\subsection{Distributions near the interface}
\label{subsec:pdfs_interface_fid}

\begin{figure*}
    \centering
    \includegraphics[width=\textwidth]{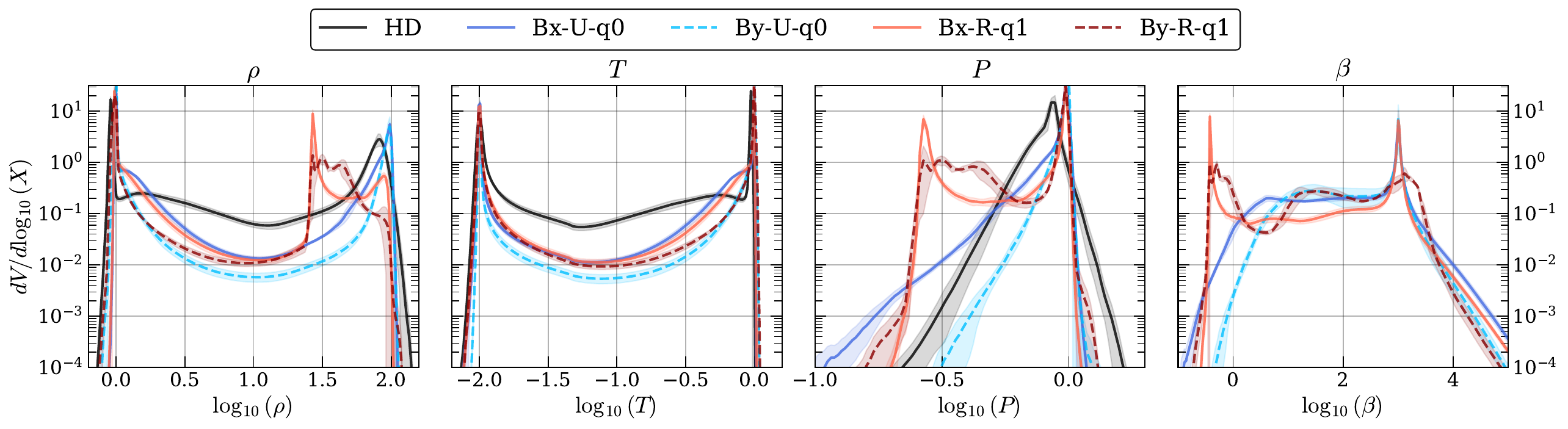}
    \caption{PDFs of density, temperature, thermal pressure and plasma beta in the interface region (\cref{eq:z_pk_region}). Shaded regions show temporal variation. MHD runs contain less intermediate-temperature gas than \texttt{HD}. Magnetically supported, thermally under-pressured cold gas forms a low-pressure tail in \texttt{Bx-U-q0} and a separate low-density peak in the polarity-reversing runs.}
    \label{fig:pdf-thermo-fid}
\end{figure*}

We next examine the probability distribution functions (PDFs) of the density, temperature, thermal pressure and plasma beta of the gas. We compute these over the regions close to the interface where gas at $T_{\rm pk}$ exists. This is given by:
\begin{align}
    Z_{\rm pk} &\equiv\{z : 1/f_w<T(x,y,z)/T_{\rm pk} < f_w\}, \label{eq:z_pk_region}\\
    z_{\rm lo} &= 0.9\left(\min(Z_{\rm pk})+ 0.5L_z\right) - 0.5L_z, \nonumber\\
    z_{\rm hi} &= 1.1\left(\max(Z_{\rm pk})+ 0.5L_z\right) - 0.5L_z, \nonumber
\end{align}
with $f_w = 10^{0.1}$. 
These distributions complement the mean vertical profiles by showing how much gas occupies each thermodynamic state, rather than only the horizontally averaged value at each height. This distinction matters for the magnetic pressure support identified in \S\ref{subsec:vert_profiles_fid}, which the mean profiles can only register as a shift of the cold-phase value.

In the \texttt{HD}, \texttt{Bx-U-q0}, and \texttt{By-U-q0} runs, the density and temperature PDFs are bimodal, with peaks corresponding to the cold and hot phases, and the pressure PDF has a single dominant peak, reflecting approximate pressure balance across the layer. Most MHD runs, however, carry substantially more gas at low thermal pressure than \texttt{HD}: at a factor of a few below the peak \texttt{Bx-U-q0} holds roughly an order of magnitude more gas, with a tail extending about a decade below it. The plasma beta PDFs show that this gas is magnetically supported. It is present in \texttt{Bx-U-q0} as well as in the runs with a strongly magnetized cold phase by construction; in \texttt{Bx-U-q0} it is produced by amplification within the layer itself but
occupies too little volume to form a separate peak.

The polarity-reversing runs show a distinct low-density, strongly magnetized population of cold gas. They also retain a broad population extending up to the larger cold-phase density of the other runs, which we associate with regions where field rearrangement or reconnection has locally removed the magnetic pressure support. A single magnetized layer therefore contains cold gas at essentially the same temperature but with densities and thermal pressures spanning a factor of a few.

The temperature PDFs show the suppression of turbulent mixing directly. Compared to \texttt{HD}, all MHD simulations contain much less gas at temperatures intermediate between the two phases, consistent with the thinner layers established in \S\ref{subsec:vert_profiles_fid} and with previous MHD TRML studies \citep{Zhao2023MNRAS}. Among the MHD runs, \texttt{By-U-q0} has the smallest intermediate-temperature gas fraction, matching its position as the run with the lowest interface area and cooling rate in \Cref{fig:time-evolution-fid}. The shapes of the temperature PDFs also differ, so the relative amount of gas at different temperatures is dependent on the presence and orientation of magnetic fields, and this could have implications for both absorption and emission properties of multiphase gas in observations \citep{Zahedy2019MNRAS,ZQu2022MNRASCUBS,Olivares2025NatAs,Chen2026arXivb}.

\subsection{Component-wise velocity and magnetic field dispersions}
\label{subsec:sigma_components_fid}

\begin{figure*}
    \centering
    \includegraphics[width=\textwidth]{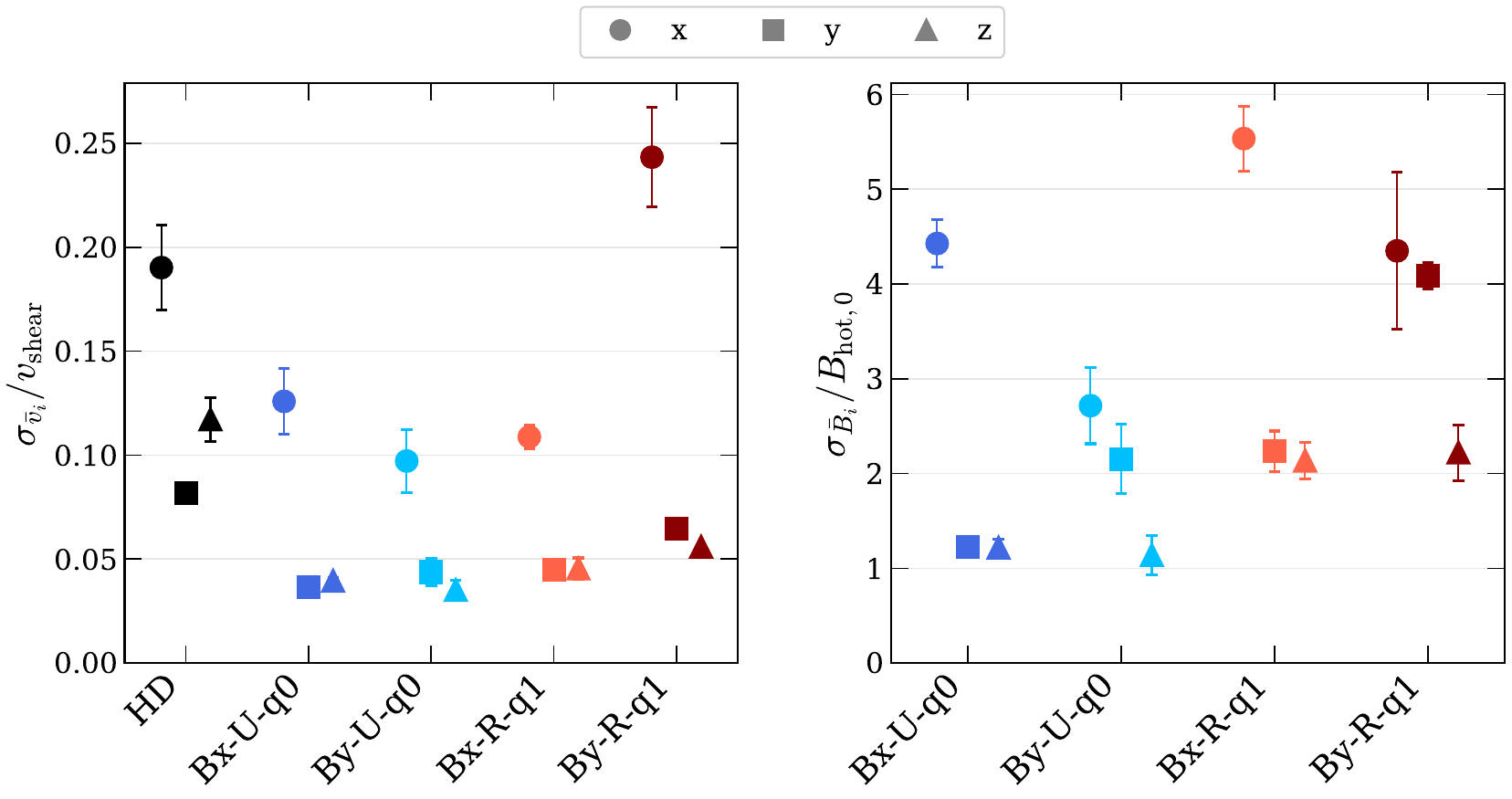}
    \caption{Dispersions of the mean-subtracted velocity (left, normalized by $v_{\rm shear}$) and magnetic field components (right, normalized by the initial hot-phase field) in the interface region. Error bars show temporal variation. $\sigma_{\bar v_x}$ dominates in every run, the transverse velocity components are suppressed in all MHD runs, and in the \texttt{By} runs $\sigma_{\bar B_x}$ grows to match or exceed $\sigma_{\bar B_y}$.}
    \label{fig:sigma-comp-fid}
\end{figure*}

In \Cref{fig:sigma-comp-fid}, we show the velocity (left) and magnetic field (right) dispersions in the same interface slab used for the PDFs above. We subtract the horizontally averaged profiles at each height, defining $\bar{v}_i\equiv v_i-\langle v_i\rangle(z)$ and $\bar{B}_i\equiv B_i-\langle B_i\rangle(z)$. The dispersions are normalized by the shear velocity and initial hot-phase field strength, respectively, and error bars show their temporal standard deviations. Corresponding second-order structure functions are presented in \S\ref{sec:appendix_sf}.

The shear-direction velocity dispersion dominates in every run and is largest in \texttt{By-R-q1}, exceeding even \texttt{HD}. Coherent streaming along its shear-aligned flux tubes (\Cref{fig:vol_rendering_fid}) contributes to this dispersion even after mean-profile subtraction. Among the remaining runs, $\sigma_{\bar{v}_x}$ is largest in \texttt{HD}.
Both $\sigma_{\bar{v}_y}$ and $\sigma_{\bar{v}_z}$ are suppressed in every MHD run relative to \texttt{HD}, and their relative sizes follow the field geometry. In the \texttt{Bx} runs, $y$ and $z$ are both perpendicular to the mean field and the two dispersions are almost equal. In the \texttt{By} runs, $y$ lies along the initial field and $\sigma_{\bar{v}_y}$ is correspondingly the larger of the two.

$\sigma_{\bar{B}_x}$ is the largest magnetic dispersion in the \texttt{Bx} runs, and is comparable to $\sigma_{\bar{B}_y}$ in the \texttt{By} runs, although their initial field has no $x$ component. In all runs the magnetic fluctuations have grown beyond the initial field strength, and the initially transverse fields develop a substantial streamwise component. This quantifies the reorientation seen in \Cref{fig:vol_rendering_fid}. The resulting field geometry may allow magnetic tension to oppose a broader range of interface motions, contributing to the stronger suppression in the \texttt{By} runs. The vertical component $\sigma_{\bar{B}_z}$ is the smallest in the \texttt{By} runs and comparable to $\sigma_{\bar{B}_y}$ in the \texttt{Bx} runs, mirroring the suppressed $\sigma_{\bar{v}_z}$. The anisotropic structure functions in \Cref{fig:sf-aniso-fid-MHD}, discussed in \S\ref{subsec:appendix_sf_aniso}, further show that both velocity and magnetic structures are elongated along the local magnetic field.

\subsection{Fractal Structure}
\label{subsec:fractal_structure_fid}

\begin{figure}
    \centering
    \includegraphics[width=0.5\textwidth]{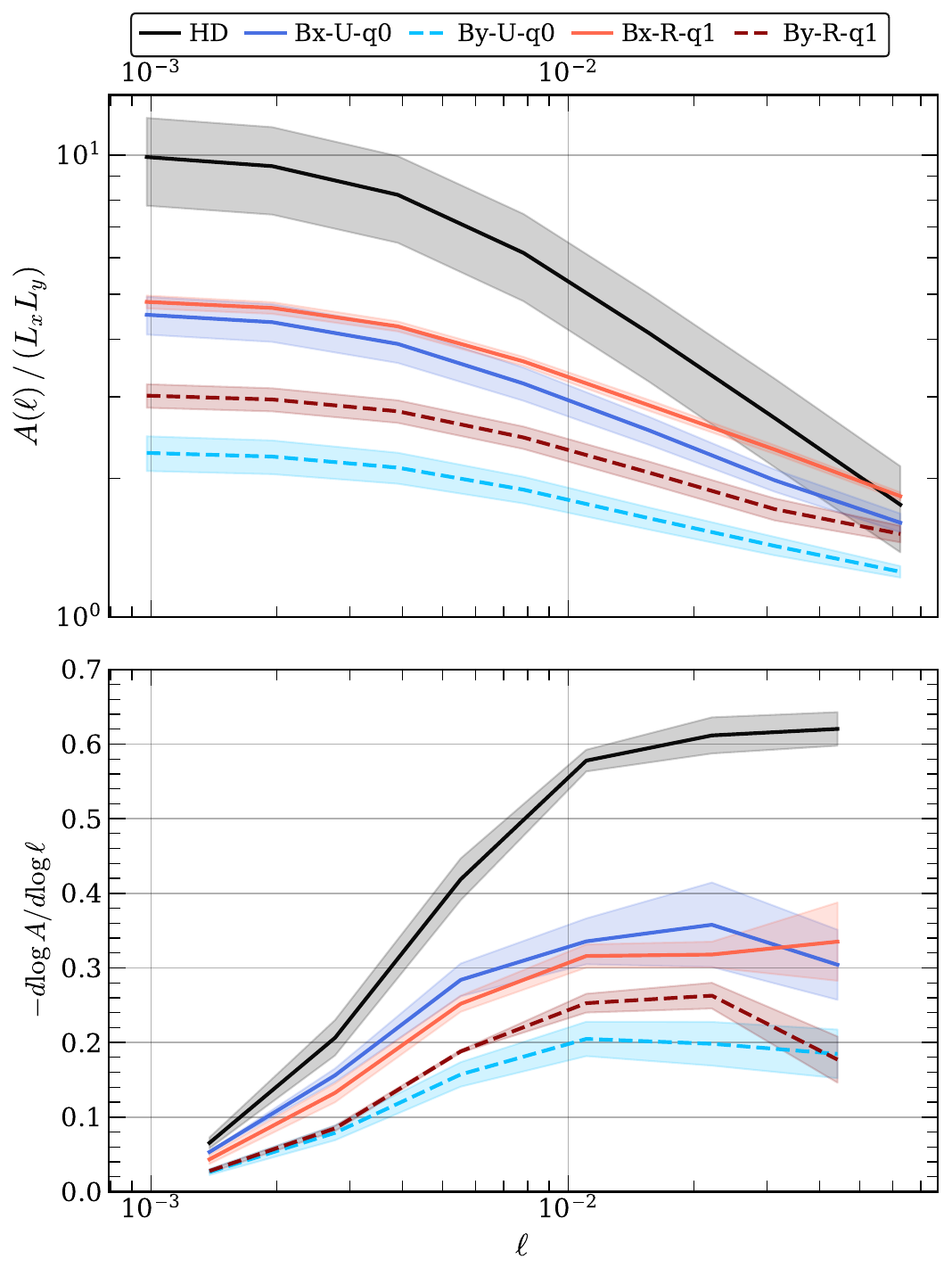}
    \caption{Scale dependence of the cooling surface area. The upper panel shows the area of the $T=T_{\rm pk}$ surface, normalized by $L_xL_y$, as a function of sampling scale, while the lower panel shows the corresponding excess fractal dimension, $d=-d\log A/d\log\ell$. The \texttt{HD} run has the largest area and the largest excess fractal dimension, $d\simeq0.6$. MHD runs have smaller areas and lower excess fractal dimensions, $d\simeq0.2$--$0.4$, showing that magnetic tension suppresses small scale folding of the cooling surface. 
    }
    \label{fig:area-vs-scale-fid}
\end{figure}

A smooth surface has a well defined area: measuring it more finely does not change its value. A fractal surface does not, because it carries structure at every scale, so each finer measurement resolves additional folds and returns a larger area. If the surface is measured at a scale $\ell$, its area grows as $A\propto\ell^{-d}$ toward small $\ell$, and the surface is said to have fractal dimension $2+d$. The quantity $d$ is the excess fractal dimension, and a smooth surface has $d=0$. Turbulent flows produce approximately fractal interfaces over the range of scales between the driving and dissipation scales \citep{Federrath2009ApJ}, and in the absence of resolved viscous scales, the hot--cold boundary of a radiative mixing layer also has a fractal-like area scaling between the inner and outer scales of the system \citep{Fielding2020ApJ}.

This matters for TRMLs because the area of the cooling surface is one of the two quantities that set the net cooling rate, together with the diffusion velocity across it \citep{Lancaster2024ApJ,Lancaster2026a}. Since the area has no single value, but depends on the measurement scale, the excess fractal dimension controls how the measured area, and hence the cooling rate, depends on numerical resolution. It is for this reason the natural quantity to compare between HD and MHD, and we use it in \S\ref{sec:diff_res} to interpret the different convergence behavior of the two.

Here, we examine the fractal structure of gas at $T\simeq T_{\mathrm{pk}}$ and how it is modified by magnetic fields. In the upper panel of \Cref{fig:area-vs-scale-fid}, we show the area of the $T=T_{\rm pk}$ surface, normalized by $L_xL_y$, measured using marching cubes at sampling scales $\ell=s\Delta x$, with $s=1,2,4,8,16,32,64$ \citep{Lewiner03_marching_cubes}. In the lower panel, we show the corresponding excess fractal dimension,
\begin{equation}
    d \equiv -\frac{d\log A}{d\log \ell},
\end{equation}
as a function of $\ell$. These quantities are time-averaged over the steady state.

As expected from the volume renderings and the velocity dispersions, the \texttt{HD} run has the largest interface area. It also has the largest excess fractal dimension, with $d\simeq0.6$, consistent with previous HD studies of radiative mixing layers \citep{Fielding2020ApJ,Lancaster2024ApJ,Lancaster2026a}. The MHD runs have systematically smaller areas, and the difference between the HD and MHD runs increases toward smaller sampling scales. This is consistent with the picture developed above: magnetic tension suppresses small scale velocity fluctuations, reducing the ability of turbulence to generate fine folds in the cooling surface.

For the MHD runs, the excess fractal dimension is smaller, with $d\simeq0.2$--$0.4$. The smallest values occur in the two runs with initially $y$ directed fields, \texttt{By-U-q0} and \texttt{By-R-q1}. In these simulations, shear also amplifies the magnetic field along the $x$ direction near the interface. This likely enhances the suppression of cooling-surface folding, reducing both the surface area and the excess fractal dimension.

Since $A\propto\ell^{-d}$, and the smallest scale available in a simulation is set by the grid, we expect the measured area to grow as $A\propto\Delta x^{-d}$ as the resolution is increased. The \texttt{HD} value $d\simeq0.6$ is therefore close to the scaling $A\propto\Delta x^{-1/2}$ that we measure directly in \S\ref{sec:diff_res}, while the smaller MHD values imply a much weaker growth of area with resolution. The excess fractal dimension measured here thus predicts, at least approximately, how the cooling rate of each run responds to increased resolution.

These results are consistent with the MHD simulations of wind blown bubbles in \citet{Lancaster2024ApJ}, where the MHD run also showed a smaller fractal dimension than the corresponding HD run. In those simulations, the reduced fractal dimension led to different convergence behavior between HD and MHD. We shall return to this issue below in \S\ref{sec:diff_res}.

\section{Dependence on cooling time and magnetization}\label{sec:results-diff-xi-mag}

We now test how the fiducial results depend on the cooling time and magnetic field strength. We vary $\xi$ to change the ratio of shear time to cooling time, vary $\beta_{\rm hot}$ to change the hot-phase magnetization, and vary $q$ to change the relative magnetization of the cold phase. Unless stated otherwise, the simulations in this section use $\Delta x=1/512$, compared to $\Delta x=1/1024$ for the fiducial set.

\subsection{Dependence on the cooling time}
\label{subsec:diff_xi}

\subsubsection{Cooling rate versus Damk\"ohler number}

\begin{figure}
    \centering
    \includegraphics[width=\columnwidth]{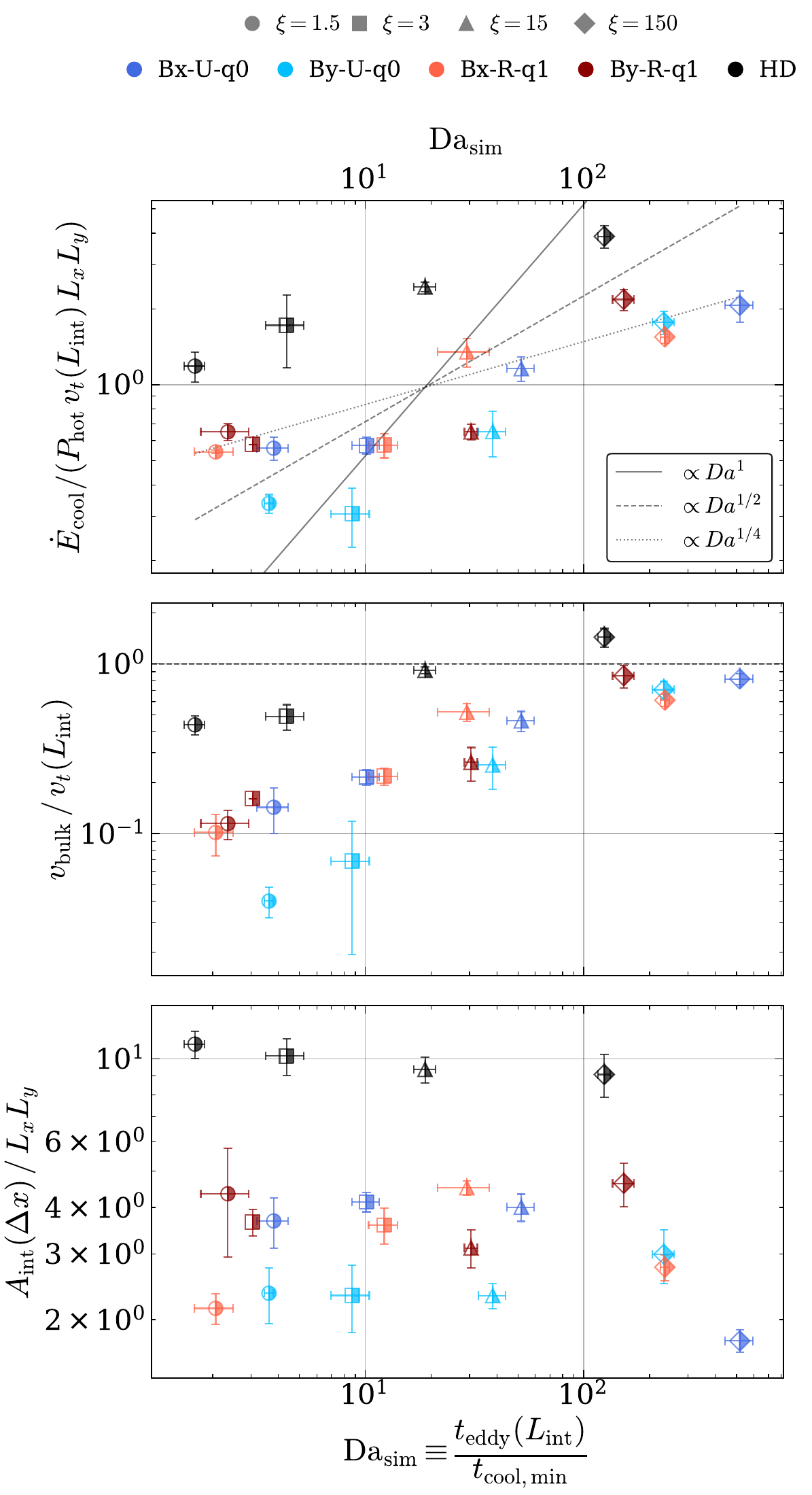}
    \caption{\emph{Upper panel:} Cooling rate normalized by the hot gas pressure $P_{\rm hot}$, the laminar area and effective turbulent velocity $v_t$ as a function of Damk\"ohler number, $\mathrm{Da}=t_{\rm eddy}/t_{\rm cool,min}$, measured on the integral scale $L_{\rm int}$. All the MHD runs show less cooling than their HD counterparts and follow a similar overall scaling with $\mathrm{Da}$, albeit with a large scatter. \emph{Middle panel:} The ratio of the hot gas bulk inflow velocity to $v_t$, which increases with increasing $\mathrm{Da}$. Inflow dominates in HD simulations for $\mathrm{Da}>100$, whereas it roughly matches the interface dispersion for the MHD runs, even at large $\mathrm{Da}$. \emph{Lower panel:} All MHD runs show a smaller interface surface area, with varying degrees of suppression.}
    \label{fig:cooling_vs_Da_diff_xi_512}
\end{figure}

In \Cref{fig:cooling_vs_Da_diff_xi_512} we show the cooling rate as a function of the Damk\"ohler number $\mathrm{Da}=t_{\rm eddy}/t_{\rm cool,min}$, measured on the integral scale of turbulence in the $Z_{\rm pk}$ interface region. The turbulent velocity $v_t$ we use is calculated from the second order structure function $\mathrm{SF}_2(\ell)$ of the mean profile-subtracted velocity components $\bar{v}_y$ and $\bar{v}_z$ (\S\ref{subsec:sigma_components_fid}). This structure function for a given field $f$ and the integral scale derived from it are given by:
\begin{subequations}
\begin{align}
    &\mathrm{SF}_2(f;\ell) = \left\langle \left| f(\mathbf{x}+\boldsymbol{\ell}) - f(\mathbf{x}) \right|^2 \right\rangle,\label{eq:structure_function}\\
    &L_{\rm int} = \frac{2}{\pi}\int_{\ell_{\min}}^{\ell_{\max}}
    \left[1-\frac{\mathrm{SF}_2(\ell)}{\mathrm{SF}_{2,\max}}\right]\,{\rm d}\ell, \label{eq:integral_scale}
\end{align}
and the effective turbulent velocity on that scale is
\begin{equation}
    v_t(L_{\rm int})=\left[\tfrac{3}{2}\left(\mathrm{SF}_2(\bar{v}_y;L_{\rm int})
        +\mathrm{SF}_2(\bar{v}_z;L_{\rm int})\right)\right]^{1/2},
\end{equation}
where the factor of $3/2$ corrects for using only the two non-shear components. The eddy turnover time on this scale is $t_{\rm eddy}=L_{\rm int}/v_t$, and $\mathrm{Da}=t_{\rm eddy}/t_{\rm cool}$, where $t_{\rm cool}$ is the minimum cooling time, which occurs at $T_{\rm pk}$ in our cooling prescription. We normalize the cooling rate by $P_{\rm hot}v_tL_xL_y$.\footnote{In contrast to
\citetalias{Lancaster2026a} and \citetalias{Lancaster2026b} we normalize by $P_{\rm hot}v_tL_xL_y$ rather than $v_t$ alone, since $P_{\rm hot}$, measured at the upper boundary, drops substantially for the larger $\xi$ runs because of our outflow boundary conditions (see \Cref{fig:vert-fid}). Our definition of $L_{\rm int}$ differs from that used in \citetalias{Lancaster2026a} and \citetalias{Lancaster2026b} who use $L_{\rm int}^\prime \equiv {\rm argmax}_{\ell} {\rm SF_2}(v_y;\ell)$. These two definitions give comparable results for HD simulations, and \Cref{fig:cool_Da_diff_norm} uses the \citetalias{Lancaster2026a} definition.}

The $\mathrm{Da}$-dependence of the cooling rate and the macro-scale bulk inflow velocity can be understood from energy flux conservation. If radiative losses are supplied predominantly by hot-gas
enthalpy, we expect
\begin{align}
    \dot{E}_{\rm cool}
    &\simeq\frac{\gamma}{\gamma-1}
    P_{\rm hot}L_xL_yv_{\rm bulk},\\
    \dot{E}_{\rm cool}
    &\sim\frac{\gamma}{\gamma-1}
    P_{\rm hot}A_{\rm int}v_{\rm diff},
\end{align}
where $A_{\rm int}\equiv A$. The second expression estimates
the enthalpy supplied to the cooling layer using its measured
area and grid-scale diffusion velocity.
Adopting $v_{\rm diff}\propto (v_t\Delta x/t_{\rm cool,min})^{1/2}$ as a model for numerical transport, motivated by the dispersion-based comparison in \S\ref{sec:appendix_enthalpy}, we obtain 
\begin{align}
    \dot{E}_{\rm cool}/(P_{\rm hot} v_t L_xL_y) &\propto A_{\rm int}/(L_xL_y)\sqrt{\mathrm{Da}\Delta x/L_{\rm int}}\text{, and} \label{eq:cooling_vs_Da_scaling}\\
    v_{\rm bulk}/v_t &\propto A_{\rm int}/(L_xL_y)\sqrt{\mathrm{Da}\Delta x/L_{\rm int}}\label{eq:vbulk_vs_Da_scaling}.
\end{align}    
\end{subequations}

In \Cref{fig:cooling_vs_Da_diff_xi_512}, we find that the \texttt{HD} runs follow $\dot{E}_{\rm cool}/(P_{\rm hot} v_t)\propto \mathrm{Da}^{0.5}$ at small $\mathrm{Da}$, with a slight flattening at large $\mathrm{Da}$, consistent with previous hydrodynamic TRML simulations \citep{Fielding2020ApJ,Tan2021MNRASa,Lancaster2026b}, and their interface area declines slowly as $\mathrm{Da}$ increases.

The MHD runs have lower cooling rates than the corresponding \texttt{HD} runs, consistent with previous MHD TRML studies \citep{Ji2019MNRAS,Zhao2023MNRAS,Das2024MNRAS}. In our simulations, the normalized cooling rates show a broadly similar dependence on $\mathrm{Da}$ in HD and MHD. The ratio of bulk to turbulent velocity and the interface area enhancement are correspondingly lower, and all three panels order the runs in the same way, so the degree to which cooling is suppressed roughly tracks the degree to which the area enhancement is suppressed. We return to this in \S\ref{subsec:disc_damkohler}.

As discussed for $\xi=15$ in \S\ref{subsec:fractal_structure_fid}, the area suppression depends on the field orientation. The shear-transverse \texttt{By-U-q0} runs are typically the most suppressed. The reconnecting runs \texttt{Bx-R-q1} and \texttt{By-R-q1} scatter more at small $\mathrm{Da}$ but follow their uniform-field counterparts at larger $\mathrm{Da}$ in net cooling. At all $\mathrm{Da}$, \texttt{By-R-q1}, which forms the flux tubes at the interface in \Cref{fig:vol_rendering_fid}, retains a larger area enhancement than the uniform-field \texttt{By-U-q0} runs.

Magnetic fields also suppress the turbulent velocity itself. Under the adopted numerical-diffusion scaling, decreasing $v_t$ at fixed interface area and integral scale moves a run along a $\mathrm{Da}^{1/2}$ relation. The vertical separation between HD and MHD therefore largely reflects differences in $A_{\rm int}$, provided that the integral scales and transport prefactors are comparable. The suppression of turbulent motions is seen more directly in \Cref{fig:time-evolution-fid} and \Cref{fig:sigma-comp-fid}. A related caveat is that $v_t$ is an imperfect tracer of
the turbulence in MHD: its $y$ component is field-aligned in the \texttt{By} runs but transverse to the field in the \texttt{Bx} runs, and its $z$ component lies along the bulk inflow, which we have attempted to remove. \citetalias{Lancaster2026a,Lancaster2026b} instead use the $y$ component of the velocity field alone; we show that version in \S\ref{sec:appendix_alt_norm}, alongside one normalized by $v_{\rm shear}$ and plotted against $\xi$, both input parameters.

\subsection{Dependence on plasma beta and cold gas magnetization}
\label{subsec:diff_magdens_magnetization}

\begin{figure}
    \centering
    \includegraphics[width=\columnwidth]{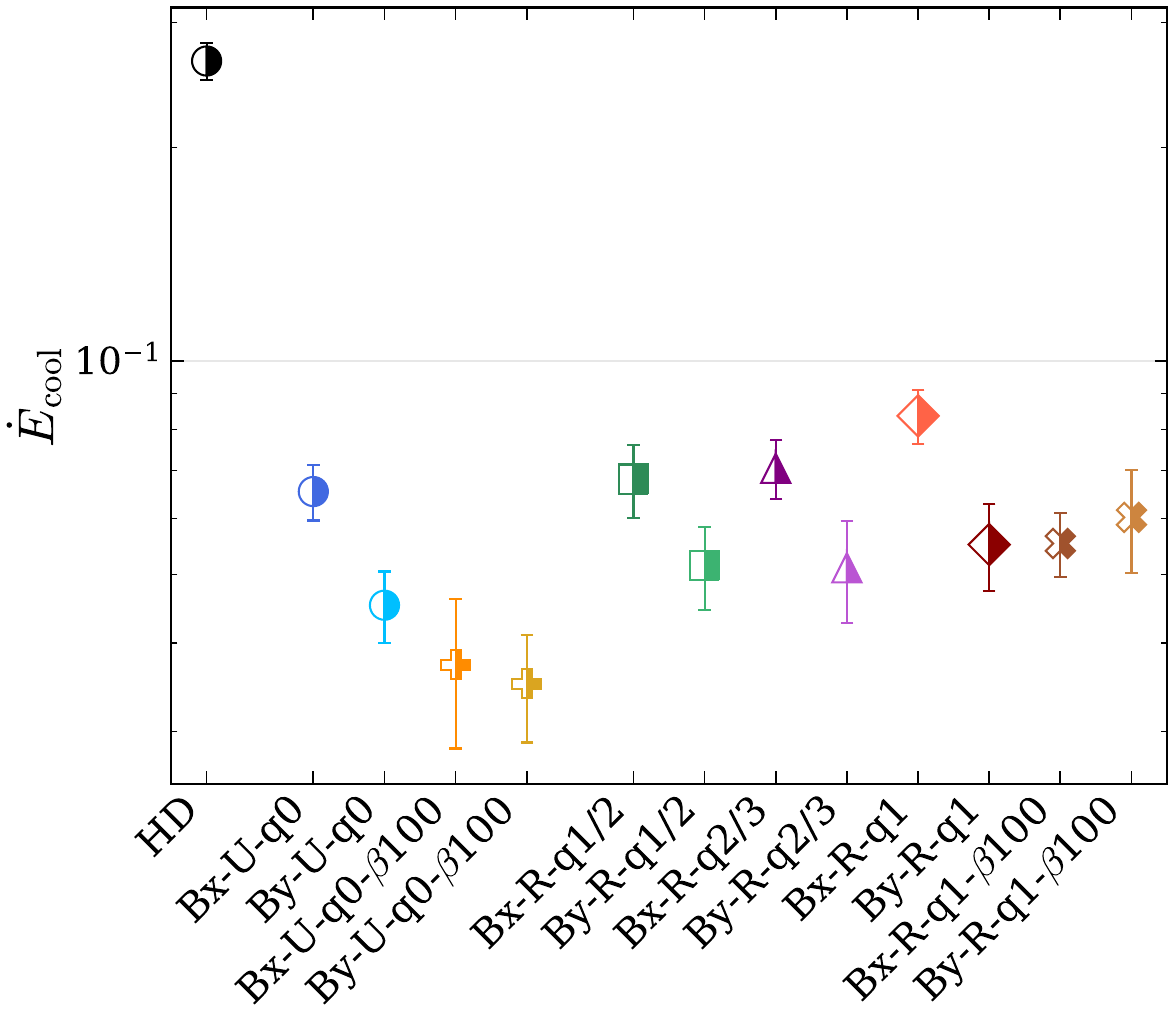}
    \caption{Cooling rate for $\xi=15$ runs with different hot-phase magnetic field strengths and different cold-phase magnetizations. Across the explored parameter range, the MHD cooling rate varies only by a factor of $\lesssim3$, much less than the overall suppression relative to HD. At fixed $\beta_{\rm hot}$ and $q$, the runs with initially transverse fields cool more slowly than their shear-aligned counterparts, with the exception of the $\beta=100$ pairs.}
    \label{fig:cooling_vs_magdens_magnetization}
\end{figure}

The fiducial MHD runs show that magnetic fields strongly suppress cooling relative to HD by reducing both the area of the cooling surface and the diffusion velocity into the mixing layer. We now ask whether this suppression depends sensitively on the initial magnetic field strength or on how strongly magnetized the cold phase is relative to the hot phase.

In \Cref{fig:cooling_vs_magdens_magnetization}, we show how the cooling rate depends on the initial magnetic field configuration for $\xi=15$ runs. Across all configurations considered here, the cooling rate varies only weakly, by a factor of $\lesssim 3$. Thus, while magnetic fields strongly suppress cooling relative to HD, the precise MHD cooling rate is only moderately sensitive to $\beta$ and $q$, within the range explored here.

One systematic trend does emerge. At $\beta_{\rm hot}=1000$ and fixed $q$, the runs with initially transverse fields cool more slowly than their shear-aligned counterparts, extending the fiducial result of \S\ref{sec:results-fid} across the parameter range. This ordering holds for every pair of configurations, but not for the $\beta_{\rm hot}=100$ pairs, suggesting that the advantage of the transverse orientation weakens once the hot phase field is strong to begin with.

We first vary the relative magnetization of the cold phase in the polarity reversing runs. The \texttt{Bx-R-q1/2} and \texttt{By-R-q1/2} runs keep $B/\rho^{1/2}$ fixed across the phases, corresponding to constant Alfv\'en speed. The \texttt{Bx-R-q2/3} and \texttt{By-R-q2/3} runs keep $B/\rho^{0.66}$ fixed, motivated by flux conservation during isotropic compression. Both sets show slightly lower cooling rates than the fiducial $q=1$ polarity reversing runs.

This weak reduction appears to be connected to the late-time magnetization of the hot--cold interface. As hot gas cools and compresses, its magnetic field is amplified in the direction opposite to the cold phase field. In the $q=1/2$ and $q=0.66$ cases, the cold phase field is weaker than in the fiducial $q=1$ case. This asymmetry may leave a stronger residual field at the interface, contributing to the slightly lower cooling rate.

We also vary the hot phase magnetic field strength by considering runs with $\beta_{\rm hot}=100$, compared to the fiducial $\beta_{\rm hot}=1000$. In general, the stronger field runs show reduced cooling relative to the fiducial field strength. They also show a smaller discrepancy between the \texttt{Bx} and \texttt{By} initial field orientations. These stronger-field runs have $\mathcal{M}_{\rm A,shear}\simeq4.6$, so the shear may be less effective at reorienting the initial field structure, reducing the extra suppression seen in the other \texttt{By} runs. Which field orientation suppresses cooling more therefore depends on $\mathcal{M}_{\rm A,shear}$.

Finally, we caution that none of the MHD cooling rates in this section is converged. These runs use $\Delta x=1/512$, and we show in \S\ref{sec:diff_res} that the MHD cooling rate continues to decrease as the resolution is increased, more rapidly for the \texttt{By} runs than for the \texttt{Bx} runs. We expect the orientation ordering described above to therefore persist, and the separation between the two field directions to widen, at higher resolution. The absolute cooling rates, and the modest differences between the $\beta_{\rm hot}$ and $q$ variations, should not be read as converged predictions.

\section{Resolution dependence}
\label{sec:diff_res}

\begin{figure*}
    \centering
    \includegraphics[width=2.0\columnwidth]{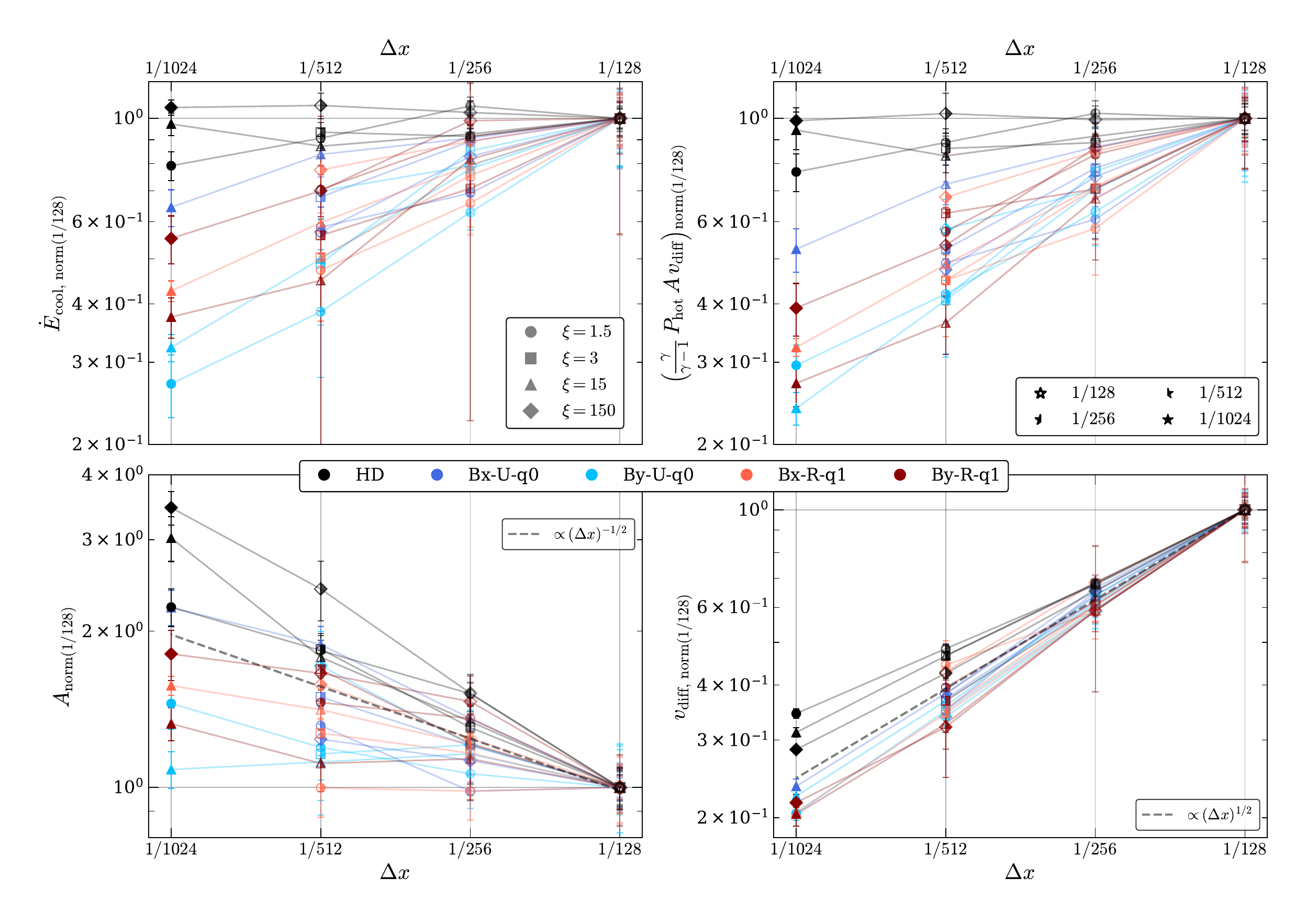}
    \caption{Resolution dependence of the net cooling rate, hot gas enthalpy flux, cooling surface area, and diffusion velocity, all normalized to their values at $\Delta x=1/128$. In HD, the cooling rate and enthalpy flux are nearly resolution independent because the cooling surface area increases as the diffusion velocity decreases. In MHD, the diffusion velocity still decreases with increasing resolution, but the cooling surface area grows too weakly to compensate because magnetic fields suppress small scale folding. As a result, the MHD cooling rate decreases with increasing resolution, and the discrepancy between HD and MHD grows at higher resolution.}
    \label{fig:cooling_vs_Area_vdiff_res}
\end{figure*}

In this section, we examine how the net cooling rate, enthalpy flux, interface area, and diffusion velocity depend on numerical resolution. In particular, we test whether the convergence of the net cooling rate reported in previous HD studies \citep{Fielding2020ApJ,Tan2021MNRASa,Lancaster2026a} and MHD simulations of \citet{Zhao2023MNRAS} also holds in our MHD simulations. We show these quantities in \Cref{fig:cooling_vs_Area_vdiff_res}, normalized in each panel by the value measured at the lowest resolution, $\Delta x=1/128$.

The net cooling rate and the enthalpy flux show similar resolution dependence, as expected from the approximate energy balance discussed above. In the HD runs, both quantities vary weakly with resolution. In the MHD runs, however, both the cooling rate and the enthalpy flux decrease with increasing resolution.

The lower panels show the origin of this difference. In the HD run, the interface area increases with resolution, approximately as $A\propto(\Delta x)^{-1/2}$, while the diffusion velocity decreases approximately as $v_{\rm diff}\propto(\Delta x)^{1/2}$. These opposing trends largely cancel in the product $A v_{\rm diff}$, yielding an approximately resolution independent enthalpy flux and cooling rate. This cancellation was reported by \citetalias{Lancaster2026a}, where the apparent convergence of the global cooling rate arose from compensating resolution dependences of the interface area and diffusion velocity.

The MHD runs behave differently. The diffusion velocity still decreases with increasing resolution, with a scaling close to $v_{\rm diff}\propto(\Delta x)^{1/2}$. However, the interface area grows much more weakly with resolution than in the HD run. This is consistent with the reduced excess fractal dimension measured for the MHD interfaces in \S\ref{subsec:fractal_structure_fid}. As a result, the decrease in $v_{\rm diff}$ is not fully compensated by an increase in area, and the net cooling rate decreases with increasing resolution.

The strength of this resolution dependence varies with magnetic field geometry. The runs with initial magnetic field along the $x$ direction show weaker suppression of the cooling rate with increasing resolution than the runs with initial field along the $y$ direction. This trend is consistent with the fractal analysis: the \texttt{Bx} runs have larger excess fractal dimensions than the \texttt{By} runs, so their interface area grows more efficiently as the resolution increases. Thus, the different convergence behavior of the MHD runs is directly connected to the suppression of small scale interface turbulence due to magnetic fields.

We therefore expect the difference between the HD and MHD runs to increase further with increasing numerical resolution, as long as dissipation is controlled primarily by numerical diffusion. We comment more on when we expect formal convergence in \S\ref{subsec:disc_convergence}.

\section{Discussion}
\label{sec:discussion}

Our primary goal in this work is to determine how magnetic fields modify the structure and energetics of turbulent radiative mixing layers. We explored uniform field configurations, similar to previous MHD TRML studies, and polarity reversing configurations in which the cold phase is more strongly magnetized and the field changes sign across the interface. These simulations test how field geometry, reconnection-prone magnetic structure, and cold phase magnetization affect mixing, cooling, and the thermodynamic structure of gas near the interface.

\subsection{How magnetic fields suppress cooling}
\label{subsec:disc_suppression}

The main result is that magnetic fields strongly suppress mixing and cooling relative to HD by reducing both turbulent transport and the geometric area available for cooling. Once magnetic tension becomes dynamically important, it resists the transverse turbulent motions that would otherwise wrinkle and broaden the cooling layer. The reduced turbulence lowers the diffusion velocity into the cooling layer, while the suppression of small scale folds reduces the excess fractal dimension of the $T_{\rm pk}$ surface. Since radiative losses are supplied primarily by hot-gas enthalpy flux, weaker turbulent transport and a smaller cooling area together lower the cooling rate.

A simple compression estimate illustrates how initially weak fields can become dynamically important as hot gas cools. For one-dimensional, flux-frozen compression perpendicular to the field, $B/B_{\rm hot}=\rho/\rho_{\rm hot}$. Assuming approximate thermal plus magnetic pressure balance with the initial hot phase gives
\begin{equation}
    \frac{\rho}{\rho_{\rm hot}}\frac{T}{T_{\rm hot}}
    +\frac{1}{\beta_{\rm hot}}\left(\frac{\rho}{\rho_{\rm hot}}\right)^2
    \simeq 1+\frac{1}{\beta_{\rm hot}}.
\end{equation}
For our fiducial $\beta_{\rm hot}=1000$, this predicts field amplification by factors of $\simeq16$ at $T_{\rm pk}$ and $\simeq27$ at $T_{\rm cold}$, corresponding to $\beta\simeq2.9$ and $\beta\simeq0.37$, respectively. The latter state also underlies our fiducial fixed-$B/\rho$ initial conditions. These estimates provide a benchmark for perpendicular compression, but do not predict a unique magnetization for the turbulent layer. The uniform-field runs develop broad plasma-beta distributions, with no distinct peak at $\beta\simeq0.37$, although \texttt{Bx-U-q0} contains a sub-unity tail (\Cref{fig:pdf-thermo-fid}).
Gas cooling through a turbulent mixing layer need not follow the one-dimensional compression assumed above, since compression along field lines can increase density without amplifying the field. Shear-driven stretching and magnetic diffusion can further modify the amplification. 

\subsection{Comparison with previous work}
\label{subsec:disc_prev_work}

This behavior is broadly consistent with previous MHD TRML studies, which found that magnetic fields are amplified in the mixing layer and reduce the amount of intermediate-temperature gas \citep{Ji2019MNRAS,Zhao2023MNRAS,Das2024MNRAS}. The amplification and magnetic back-reaction are also consistent with nonlinear MHD KH studies \citep{Frank1996ApJ,Ryu2000ApJ} and the broader small scale dynamo picture \citep[e.g.][]{Schekochihin2004ApJ,Seta2021PhRvF,Beattie2023MNRAS}.

The field orientation dependence is one of our more striking results, and it runs opposite to the expectation from linear theory. A shear-aligned field is bent by the growing KH modes and its tension opposes them, while a transverse field satisfies $\mathbf{k}\cdot\mathbf{B}=0$ for a perturbation in the plane of the shear and provides no restoring tension. On this basis the \texttt{Bx} runs should suppress mixing more effectively than the \texttt{By} runs, but we find the reverse. A possible explanation is that the layer does not remain two-dimensional: once turbulence develops, the super-Alfv\'enic shear reorients part of the initially transverse field toward the streamwise direction, and \S\ref{subsec:sigma_components_fid} shows that in both \texttt{By} runs the fluctuating field ends up with comparable dispersions along $x$ and $y$. The resulting field geometry may allow magnetic tension to oppose a broader range of interface motions, contributing to the stronger suppression. This effect of the field orientation is observed in most of our simulations, except our more strongly magnetized $\beta_{\rm hot}=100$ runs, where $\mathcal{M}_{\rm A,shear}\simeq4.6$.

This may help explain why our result differs from earlier work. \citet{Das2024MNRAS} found weaker suppression in their transverse-field TRML runs than in their shear-aligned cases, which is what linear theory predicts, while \citet{Ji2019MNRAS} found only a weak dependence on the initial field orientation. Differences in resolution and run duration may contribute to this discrepancy by affecting how far field amplification proceeds (\S\ref{subsec:time_evol_fid}). Those studies used $\Delta x\simeq1/64$--$1/128$, whereas our fiducial runs reach $\Delta x=1/1024$. The reorientation that drives the effect requires the interface turbulence to be resolved well enough for turbulence-driven amplification to proceed, and the sensitivity of small-scale dynamo action to numerical resolution is well established \citep[e.g.][]{Schekochihin2004ApJ,Federrath2011ApJ}. Our own resolution study points the same way: the \texttt{By} runs have the smallest excess fractal dimensions and show the strongest decrease in cooling rate with increasing resolution.

A further difference concerns the relation between turbulence and cooling. \citet{Das2024MNRAS} found that the HD relation between turbulent velocity and cooling rate also describes their MHD simulations, and argued that magnetic fields affect cooling only through their influence on turbulence generation. Our results likewise link the magnetic suppression of cooling to weaker turbulence, but indicate that a single turbulent velocity amplitude does not fully describe this effect. Although the HD and MHD runs show broadly similar dependencies on $\mathrm{Da}$, normalizing the cooling rate by $P_{\rm hot}v_tL_xL_y$ does not collapse them onto a common relation: the MHD runs retain lower normalized cooling rates at comparable $\mathrm{Da}$ (\Cref{fig:cooling_vs_Da_diff_xi_512}). This residual suppression accompanies a smaller cooling-surface area and a lower excess fractal dimension (\Cref{fig:area-vs-scale-fid}). Thus, magnetic fields modify how turbulent motions fold the interface and create area for mixing, in addition to reducing their amplitude. Our results indicate that a single measure of turbulent velocity does not fully capture these geometric effects on the cooling rate.
\subsection{Polarity reversal and reconnection}
\label{subsec:disc_reconnection}

Although reconnection could reduce magnetic tension, the polarity reversing configurations show different cooling-gas morphologies, including flux-tube-like structures, without restoring HD-like mixing. Their cooling rates follow the same relation with hot gas enthalpy flux as the other simulations. Thus, for the initially weak hot-phase fields considered here, reconnection heating does not appear to dominate the global energy budget in our ideal MHD calculations. Reconnection may nevertheless facilitate mixing by relaxing the magnetic tension that opposes motions across the interface, a possibility we return to in \S\ref{subsec:disc_convergence}. Its role may depend on the initial magnetization and on the treatment of resistive dissipation.

\subsection{Dependence on the Damk\"ohler number}
\label{subsec:disc_damkohler}

The broadly similar dependence of the normalized cooling rate on Damk\"ohler number in HD and MHD is consistent with their common numerical-transport scaling (\Cref{fig:vdiff_vs_vturb}). By suppressing the turbulent motions that fold the interface, magnetic fields reduce its area by an amount that changes little as the cooling time is varied, contributing to the lower normalization of this relation (lower panel of \Cref{fig:cooling_vs_Da_diff_xi_512}). The shear-transverse runs typically show the strongest area suppression for $\xi\leq15$. 

The slight flattening of the \texttt{HD} runs at the largest $\mathrm{Da}$ has a separate origin: there the interface area itself declines slowly as cooling becomes faster (lower panel of \Cref{fig:cooling_vs_Da_diff_xi_512}, discussed in \citetalias{Lancaster2026b}), bending the curve without requiring a change in the underlying transport scaling.

\subsection{Numerical convergence}
\label{subsec:disc_convergence}

The resolution study highlights an important numerical consequence of this geometric suppression. In HD TRMLs, the net cooling rate can appear nearly converged because two resolution dependencies compensate: the cooling surface area increases with resolution, while the diffusion velocity decreases \citepalias{Lancaster2026a}. In MHD, the diffusion velocity still decreases with increasing resolution, but the cooling surface area grows much more weakly because magnetic fields suppress small scale folding. The HD cancellation therefore no longer operates in the same way.

The two quantities behave differently at our highest resolutions. 
Between $\Delta x=1/512$ and $1/1024$ the MHD interface areas change relatively little, typically by about 20 per cent or less, consistent with their small excess fractal dimensions, whereas the HD areas grow by 20--70 per cent. The diffusion velocity keeps falling in every run, as expected when it is set by numerical transport (\S\ref{subsec:time_evol_fid}). With the area growing slowly and $v_{\rm diff}$ still falling, the MHD cooling rate shows no sign of converging at the resolutions we reach.

One possible route to convergence is for physical thermal conduction to dominate numerical heat diffusion across the interface and thereby set $v_{\rm diff}$ independently of $\Delta x$. In that regime, $v_{\rm diff}$ would need to be evaluated across the resolved conductive layer rather than over the grid-scale thickness used here. In a conduction-dominated front, the characteristic thermal transition scale is the conductive Field length, where conduction balances radiative cooling; resolving this scale would allow the conductive front structure and its associated $v_{\rm diff}$ to be captured \citep{Koyama2004ApJ,Tan2021MNRASa}. In a magnetized layer, however, conduction occurs primarily along magnetic field lines, so the relevant Field length depends on the conductivity normal to the local interface. Neglecting conduction perpendicular to the magnetic field, we write
\begin{subequations}\label{eq:conductive_transport}
\begin{align}
    \boldsymbol{q}_{\rm cond}
    &= -\kappa_\parallel\hat{\boldsymbol{b}}
    \left(\hat{\boldsymbol{b}}\cdot\boldsymbol{\nabla}T\right),
    \label{eq:conductive_heat_flux}\\
    \kappa_n
    &= \kappa_\parallel
    \left(\hat{\boldsymbol{b}}\cdot\hat{\boldsymbol{n}}\right)^2,
    \label{eq:normal_conductivity}\\
    \lambda_{{\rm F},n}(P,T_{\rm pk})
    &\sim
    \left[
    \frac{\kappa_n(T_{\rm pk})T_{\rm pk}}
    {\mathcal{L}_{\rm max}(P)}
    \right]^{1/2}
    \notag\\
    &=
    \left[
    \frac{(\gamma-1)\kappa_n(T_{\rm pk})T_{\rm pk}
    t_{\rm cool,min}}
    {\bar{P}}
    \right]^{1/2}
    \frac{\bar{P}}{P},
    \label{eq:normal_field_length}
\end{align}
\end{subequations}
where $\kappa_\parallel$ is the field-aligned conductivity, $\hat{\boldsymbol{b}}$ is the magnetic-field direction, and $\hat{\boldsymbol{n}}$ is defined in \cref{eq:isosurface_normal}. We evaluate the Field length at $T_{\rm pk}$ using \cref{eq:cooling_rate_max}, neglecting the small heating correction there. We find that the fields preferentially align tangent to the interface, with $\langle(\hat{\boldsymbol{b}}\cdot\hat{\boldsymbol{n}})^2\rangle<0.05$ in all MHD simulations in the fiducial set. This geometry strongly suppresses the normal conductivity and reduces the corresponding Field length. Consistent with this expectation, \citet{Zhao2023MNRAS} found that constant, Spitzer, and reduced Spitzer conductivity prescriptions produced very similar cooling rates in their magnetized layers, with only minor changes to the temperature distributions, primarily in the hot gas. Whether the residual conductive transport can set $v_{\rm diff}$ and yield a converged cooling rate in our simulations requires explicit tests with anisotropic conduction. 

A second possibility is a transition to plasmoid-mediated reconnection as increasing resolution reduces numerical resistivity and raises the effective Lundquist number of the current sheets above the plasmoid-instability threshold. The resulting tearing-mediated, plasmoid-dominated reconnection regime has a rate that is only weakly dependent on the global resistivity, potentially saturating the observed scaling if reconnection controls the transport into the cooling layer \citep{Bhattacharjee2009PhPl,Ripperda2019MNRAS,Fielding2023ApJ}. 

A complementary criterion for resolving the thermal structure is that of \citetalias{Lancaster2026a}, who argue that the turbulent Field length $\lambda_{\rm F,turb}$, the scale at which the eddy turnover time equals the cooling time, must be well resolved by the grid. This is a more demanding requirement in MHD than in HD because the turbulent velocities are suppressed overall, which pushes $\lambda_{\rm F,turb}$ to smaller scales at fixed $t_{\rm cool}$ (\Cref{fig:sf-fid}). Resolving the magnetized case is therefore harder than the hydrodynamic one, even before the additional dissipative scales introduced by the field are considered. If explicit conduction is included, a local effective-diffusion picture suggests that the thermal transition scale is set approximately by the larger of $\lambda_{\rm F,turb}$ and $\lambda_{{\rm F},n}$.

\subsection{Implications for observations}
\label{subsec:disc_observations}

Recent work has emphasized that the geometry of multiphase gas can strongly affect temperature PDFs and emission diagnostics \citep{ZChen2026OJAp,Chen2026arXivb}. Our simulations add a related caveat: magnetic fields can modify the geometry itself. By suppressing small scale folding of the cooling surface and reducing the excess fractal dimension, MHD changes both the amount of intermediate-temperature gas and the rate at which hot gas is transported into the cooling layer. The changes in the temperature distributions in \Cref{fig:pdf-thermo-fid} could alter the relative H$\alpha$ and X-ray emission, although quantitative predictions require emissivity-weighted calculations and simulations run with the appropriate cooling curve. Therefore, it is unclear whether geometric relations found in HD, including the X-ray to H$\alpha$ surface brightness relation in \citet{Chen2026arXivb}, will carry over unchanged to magnetized mixing layers.

Although this paper is primarily theoretical, the results also suggest a caution for interpreting thermal pressures inferred from CGM absorption models. Absorption line analyses use ionic column densities and line widths, together with ionization modeling, to infer the density, temperature, and thermal pressure of cool CGM gas \citep{Zahedy2019MNRAS,ZQu2022MNRASCUBS,ZQu2023MNRASCUBS}. These inferred pressures are often compared to expectations from pressure balance with other cool components or with the ambient hot halo. Our simulations show that local thermal pressure balance need not hold if cold gas is magnetically supported. A weakly magnetized hot phase can coexist with a strongly magnetized cold phase, so cold gas may appear substantially thermally under-pressured (by $\gtrsim50\%$) even when the total pressure is balanced. In reconnecting configurations, local loss of magnetic pressure support can also produce denser cold gas.


\section{Conclusions}
\label{sec:Conclusion}

We have presented a suite of three dimensional MHD simulations of turbulent radiative mixing layers to study how magnetic field geometry, cooling strength, and numerical resolution affect mixing and cooling. In addition to uniform fields along and transverse to the shear direction, we consider polarity reversing configurations, where the magnetic field in the cold phase is oppositely directed relative to the field in the hot phase, producing current sheets and reconnection-prone structures near the interface. Our main conclusions are as follows.

\begin{itemize}

    \item Magnetic fields reduce turbulent mixing and radiative cooling relative to HD simulations. The suppression occurs because magnetic tension weakens the transverse turbulent motions that transport hot gas into the cooling layer and suppresses the small-scale wrinkling of the cooling surface. As a result, the MHD runs have smaller cooling-surface areas and smaller diffusion velocities, which together reduce the hot-gas enthalpy supply and hence the net cooling rate (\Cref{fig:time-evolution-fid,fig:area-vs-scale-fid,fig:sigma-comp-fid}).

    \item The orientation of the magnetic field has a secondary effect. In the super-Alfv\'enic flows considered here, fields initially aligned with the shear suppress cooling less strongly than fields initially transverse to the shear. In the transverse field runs, shear bends part of the field toward the flow direction, giving the field comparable components along and across the shear near the interface which contributes to the stronger suppression of turbulence and folding of the interface (\Cref{fig:time-evolution-fid,fig:sigma-comp-fid}). In our more strongly magnetized runs ($\beta_{\rm hot}=100$, $\mathcal{M}_{\rm A,shear}\simeq4.6$), the effect of field orientation is minor.

    \item Polarity reversal changes the morphology of the cooling gas but does not restore HD-like mixing. Although oppositely directed fields create current sheets, reconnection-prone structures, and flux tube like morphology, the cooling rate remains suppressed. For the flows considered here, which are subsonic and super-Alfv\'enic in the hot phase, the cooling rates remain consistent with radiative losses being supplied primarily by hot-gas enthalpy, without evidence for an additional dominant heating contribution from reconnection (\Cref{fig:vol_rendering_fid,fig:cooling_vs_Area_vdiff}).

    \item Across the cooling times explored, all MHD configurations show suppressed interface areas and cooling rates relative to HD. The shear-transverse runs typically show stronger suppression. The normalized cooling rates have a broadly similar dependence on Damk\"ohler number in HD and MHD (\Cref{fig:cooling_vs_Da_diff_xi_512,fig:cooling_vs_magdens_magnetization}).

    \item HD and MHD show different apparent convergence behavior. In HD, increasing resolution increases the cooling surface area while decreasing the diffusion velocity, leading to an approximate cancellation in the net cooling rate. In MHD, magnetic fields suppress the growth of cooling surface area, so this cancellation is weakened. At higher resolution, faster dynamo amplification also leads to earlier suppression of turbulent motions. The MHD cooling rate therefore decreases with increasing resolution, and the difference between HD and MHD grows at higher resolution (\Cref{fig:cooling_vs_Area_vdiff_res,fig:cool_rate_uni_diff_res}).

    \item Cold gas in magnetized mixing layers need not be in local thermal pressure balance with the hot phase. Magnetic pressure support can lower the thermal pressure and density of cold gas, while local loss of magnetic pressure support in polarity reversing runs can produce denser cold gas. Thus, a single magnetized mixing layer can contain cold gas at similar temperatures but with different densities and thermal pressures (\Cref{fig:pdf-thermo-fid}).

    \item Quantitative predictions from ideal MHD simulations should be interpreted with caution. In our simulations, the diffusion velocity and cooling rate remain resolution dependent. Converged predictions of cooling rates, temperature distributions, and emission require establishing that the transport into the cooling layer is independent of numerical resolution.

    \item Achieving formal convergence may remain difficult even with explicit thermal conduction. The predominantly tangential fields at the interface, with $\langle(\hat{\boldsymbol b}\cdot\hat{\boldsymbol n})^2\rangle<0.05$ in our fiducial MHD runs, would strongly suppress the conductivity normal to the interface and reduce the corresponding Field length. This smaller scale must be resolved for conduction to set $v_{\rm diff}$ independently of the grid, making this route to convergence substantially more demanding than with isotropic conduction.

\end{itemize}

\section*{Acknowledgements}
RM thanks Eve C.Ostriker for useful discussions. The authors gratefully acknowledge the support of the Kavli Institute for Theoretical Physics's 2024 program on ``Turbulence in Astrophysical Environments,'' where this work was initially conceived, and that of the Aspen Center for Physics's program ``Toward a Holistic Understanding of the Multi-scale, Multiphase Circumgalactic Medium,'' where this work was continued. This research was therefore supported in part by grant NSF PHY-2309135 to the Kavli Institute for Theoretical Physics and NSF PHY-2210452 to the Aspen Center for Physics. This work was supported by National Science Foundation (NSF) grants AST-2107872 and AST-2509269 (PI: Quataert). L.L. acknowledges the support of the Simons Foundation under grant 965367.
DBF gratefully acknowledges support from NSF through grant AST-2407387 and from NASA through grants HST-AR-17859.015-A and HST-AR-17559.009-A. This work was supported by a grant from the Simons Foundation (Grant Award ID BD-Targeted-00017375, DBF).
GLB acknowledges support from the NSF (AST-2307419) and NASA (80NSSC21K1053), as well as support from the Simons Foundation through the Learning the Universe Collaboration.

The analysis presented in this article was performed in part on computational resources managed and supported by Princeton Research Computing, a consortium of groups including the Princeton Institute for Computational Science and Engineering (PICSciE) and the Office of Information Technology's High Performance Computing Center and Visualization Laboratory at Princeton University.

This research used both the DeltaAI advanced computing and data resource, which is supported by the National Science Foundation (award OAC 2320345) and the State of Illinois, and the Delta advanced computing and data resource, which is supported by the National Science Foundation (award OAC 2005572) and the State of Illinois. Delta and DeltaAI are joint efforts of the University of Illinois Urbana-Champaign and its National Center for Supercomputing Applications. This work used the Delta system at the National Center for Supercomputing Applications through allocations PHY230106 and PHY230045 from the Advanced Cyberinfrastructure Coordination Ecosystem: Services \& Support (ACCESS) program, which is supported by National Science Foundation grants 2138259, 2138286, 2138307, 2137603, and 2138296.

An award of computer time was provided by the U.S. Department of Energy's Innovative and Novel Computational Impact on Theory and Experiment (INCITE) Program under allocation AST238 (PI: Quataert). The simulations performed in this work used resources of the Oak Ridge Leadership Computing Facility at Oak Ridge National Laboratory, which is supported by the Office of Science of the U.S. Department of Energy under Contract No. DE-AC05-00OR22725.

RM used the large language models Claude (Anthropic; Sonnet 5 and Opus 5) and ChatGPT (OpenAI; GPT 5.5 and Astra 6) to assist with manuscript revision and the development of simulation analysis scripts. All AI-assisted text and code were reviewed and verified by the authors, who take full responsibility for the manuscript.

\section*{Software} 
{\texttt{AthenaK} \citep{Stone2020ApJS,Stone2026ApJS}, \texttt{matplotlib} \citep{Hunter4160265}, \texttt{cmasher} \citep{Ellert2020JOSS}, \texttt{scipy} \citep{Virtanen2020}, \texttt{NumPy} \citep{Harris2020}, \texttt{CuPy} \citep{Okuta2017CuPyA}, \texttt{h5py} \citep{collette_python_hdf5_2014}, and \texttt{astropy} \citep{astropy2018}}.


\section*{Data Availability}
All relevant data associated with this article are available upon reasonable request to the corresponding author.

\section*{Additional Links}
Movies of our simulations are available as online supplementary material.


\bibliographystyle{mnras}
\bibliography{refs.bib} 



\appendix

\setcounter{section}{0}
\setcounter{figure}{0}
\renewcommand{\thesection}{\Alph{section}}
\renewcommand{\thefigure}{\thesection\arabic{figure}}
\makeatletter
\@addtoreset{figure}{section}
\makeatother

\section{Other scale-dependent statistics}
\label{sec:appendix_sf}

\subsection{Component-wise structure functions}
\label{subsec:appendix_sf_comp}

\begin{figure*}
    \centering
    \includegraphics[width=\textwidth]{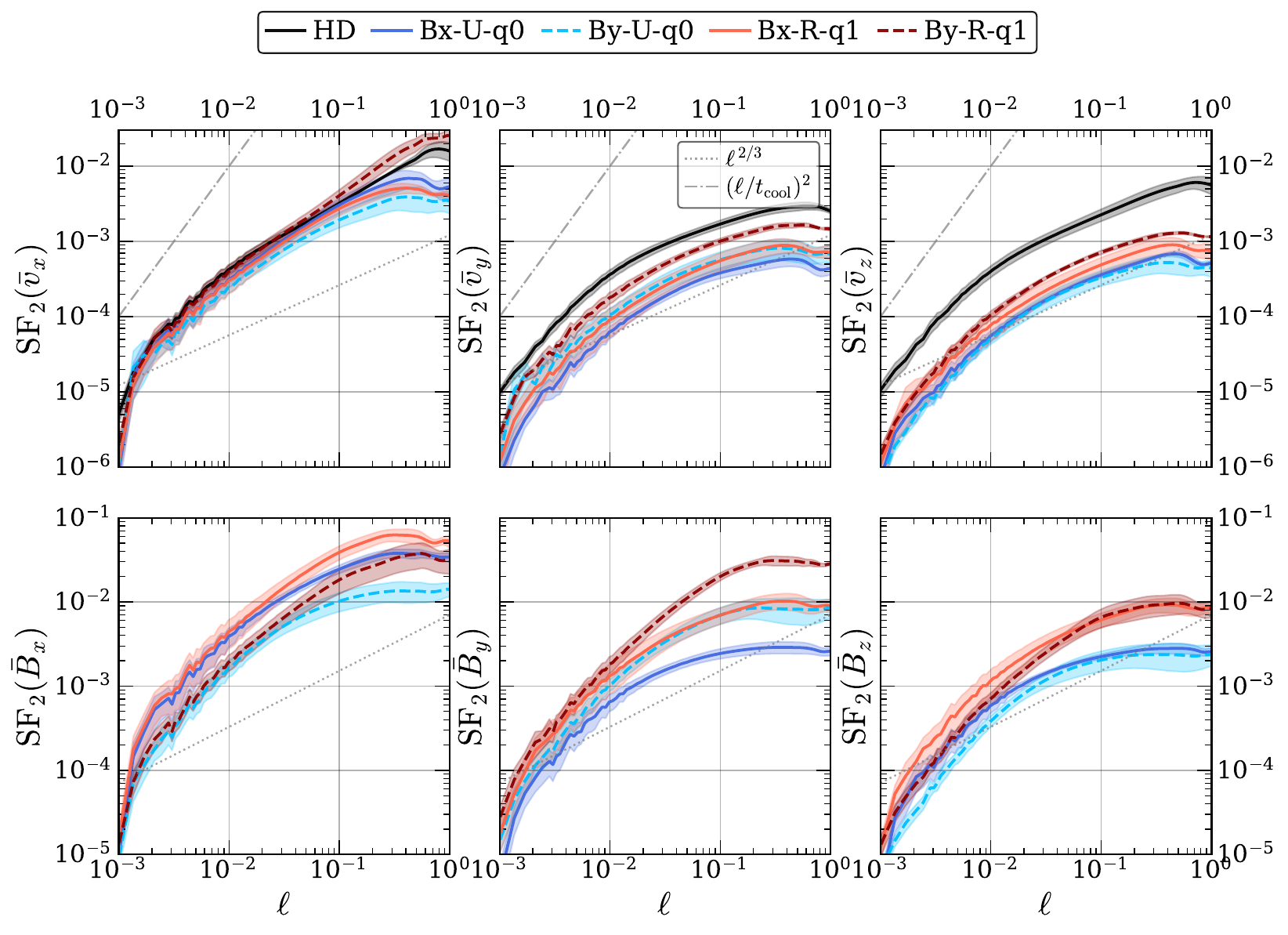}
    \caption{Second order structure functions of the mean profile subtracted velocity components $\bar{v}_i$ (top row) and magnetic field components $\bar{B}_i$ (bottom row), measured in the interface region for the fiducial runs. The relative amplitudes broadly follow those of the dispersions in \Cref{fig:sigma-comp-fid}, and the suppression of non-shear motions in MHD extends across the resolved scales.}
    \label{fig:sf-fid}
\end{figure*}

The dispersions in \S\ref{subsec:sigma_components_fid} integrate over all scales in the interface slab. Here we check that the differences between runs are not confined to a particular range of separations. \Cref{fig:sf-fid} shows $\mathrm{SF}_2$, defined in \Cref{eq:structure_function}, for the three velocity components in the top row and for the three magnetic field components in the bottom row.

The $\mathrm{SF}_2$ do not show an extended inertial range with the canonical $\ell^{2/3}$ scaling expected for homogeneous incompressible turbulence \citep{kolmogorov1941dissipation}. This is not surprising: the flow is highly inhomogeneous, radiatively cooling, and dominated by a sharp hot--cold boundary, so the assumptions behind Kolmogorov scaling do not apply. Similarly shallow $\mathrm{SF}_2$ were found in \citetalias{Lancaster2026a,Lancaster2026b}. The finite dynamic range of the simulations and the bottleneck effect can also modify the apparent slope near the grid scale relative to the asymptotic inertial-range expectation \citep{Ishihara2016PhRvF,Yeung2025JFM}. We therefore compare amplitudes rather than fitting power-law slopes, which is the reason the main text uses the component dispersions instead.

The relative amplitudes are similar to those of the dispersions in \Cref{fig:sigma-comp-fid}. The shear-aligned component $\mathrm{SF}_2(\bar{v}_x)$ is the largest for the \texttt{By-R-q1} and \texttt{HD} runs, but for the former, the large value could be partly due to the insufficient removal of laminar shear motions through our mean profile subtraction. The non-shear velocity components are smaller in the MHD runs than in \texttt{HD}. The \texttt{Bx} and \texttt{By} runs also separate in the same sense as in \Cref{fig:sigma-comp-fid}. These differences persist across the full range of $\ell$, showing that magnetic fields suppress non-shear motions at all resolved scales, rather than only near the grid scale.

\subsection{Anisotropic structure functions}
\label{subsec:appendix_sf_aniso}

\begin{figure*}
    \centering
    \includegraphics[width=\textwidth]{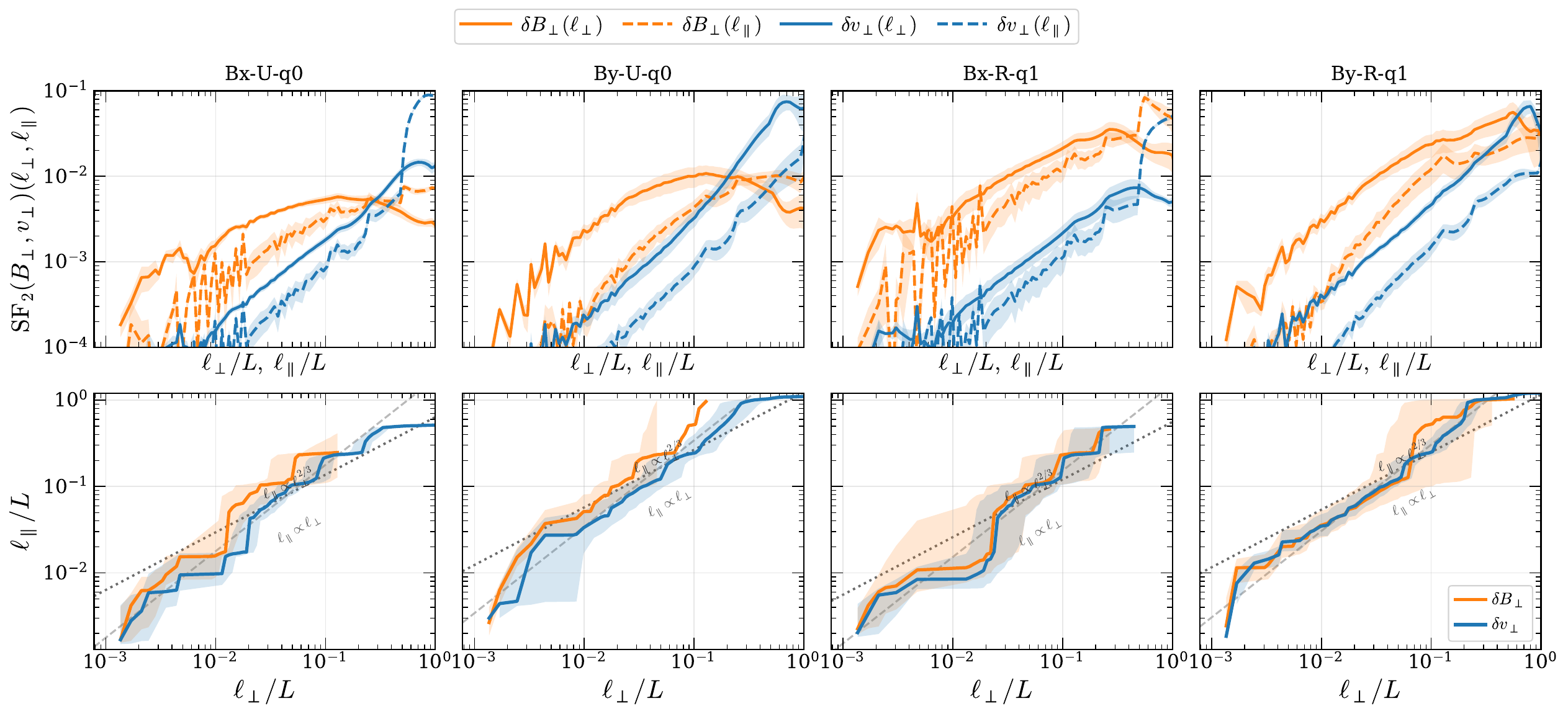}
    \caption{\emph{Top row:} Anisotropic second order structure functions of the perpendicular velocity and magnetic field increments, measured as a function of separations parallel and perpendicular to the local mean magnetic field. In all runs, fluctuations are larger for perpendicular separations, showing that velocity and magnetic structures are elongated along the local field. \emph{Bottom row:} Inferred anisotropy scale, defined as the parallel separation at which the parallel and perpendicular structure functions become equal for a given perpendicular separation. The scaling is closer to $\ell_\parallel\propto\ell_\perp$ than to the \citetalias{Goldreich1995ApJ} expectation.}
    \label{fig:sf-aniso-fid-MHD}
\end{figure*}

We next measure how the structure functions depend on the direction of the separation vector relative to the local mean magnetic field, following a procedure similar to \citet{Fielding2023ApJ}. For a pair of points separated by $\boldsymbol{\ell}$, we define the local mean field as $\mathbf{B}_{\ell}\equiv[\mathbf{B}(\mathbf{x})+\mathbf{B}(\mathbf{x}+\boldsymbol{\ell})]/2$, and compute the angle
\begin{equation}
    \theta_{\ell} =
    \cos^{-1}\left(
    \frac{|\mathbf{B}_{\ell}\cdot\boldsymbol{\ell}|}
    {|\mathbf{B}_{\ell}|\,|\boldsymbol{\ell}|}
    \right).
\end{equation}
We classify point pairs as parallel when $0\leq\theta_\ell<\pi/18$ and perpendicular when $4\pi/9\leq\theta_\ell<\pi/2$, and in each separation bin compute $\mathrm{SF}_2$ of the velocity and magnetic field components perpendicular to the local mean field, $\delta v_\perp$ and $\delta B_\perp$.

The top row of \Cref{fig:sf-aniso-fid-MHD} shows that in all four MHD runs, both velocity and magnetic fluctuations are larger for perpendicular than for parallel separations over most of the resolved range. Fluctuations therefore decorrelate more rapidly across the local field than along it, and both velocity and magnetic structures are elongated along the local mean field.

The bottom row shows the inferred anisotropy scale: for each $\ell_\perp$, the value of $\ell_\parallel$ at which the parallel and perpendicular structure functions are equal, computed separately for $\delta v_\perp$ and $\delta B_\perp$. In contrast to \citet{Fielding2023ApJ}, who found a \cite{Goldreich1995ApJ} (hereafter \citetalias{Goldreich1995ApJ})-like scaling, $\ell_\parallel\propto\ell_\perp^{2/3}$, our simulations show a relation closer to $\ell_\parallel\propto\ell_\perp$. The flattening at the largest separations is likely due to poorer sampling and the finite thickness of the interface rather than a robust change in scaling. The difference from \citet{Fielding2023ApJ} may reflect our weaker initial magnetization: their simulations start at $\beta=1$, whereas our uniform-field runs begin at $\beta_{\rm hot}=1000$ with $\mathcal{M}_{\rm A,shear}\sim14$. A \citetalias{Goldreich1995ApJ}-like cascade is expected only over scales where magnetic tension is dynamically important and the flow is not yet dominated by numerical dissipation.

\section{Cooling rate versus enthalpy flux}
\label{sec:appendix_enthalpy}

\begin{figure}
    \centering
    \includegraphics[width=\columnwidth]{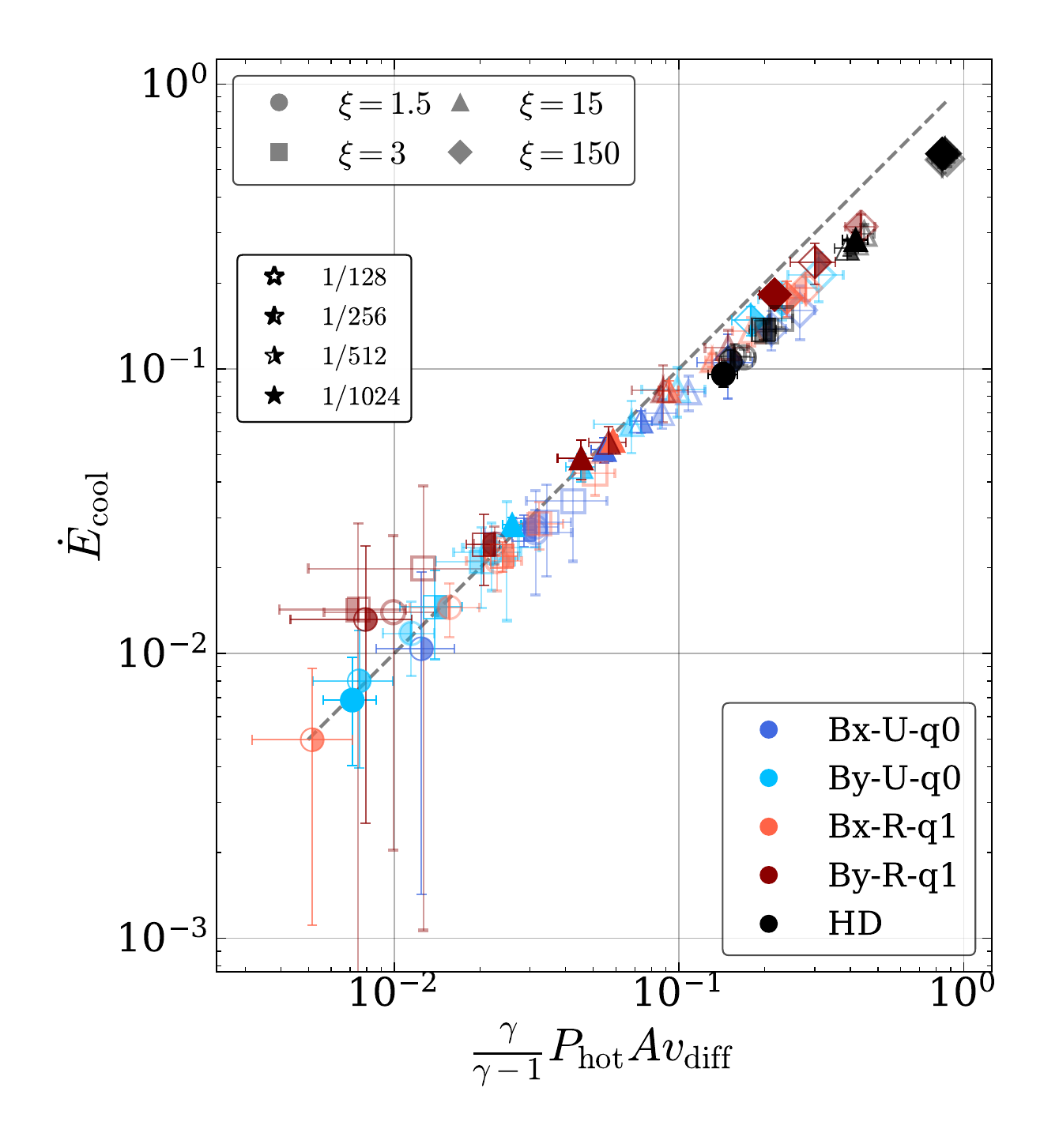}
    \caption{Net cooling rate versus the estimated enthalpy supply $\gamma P_{\rm hot}A v_{\rm diff}/(\gamma-1)$ for runs with different cooling times and magnetic configurations. Both quantities are divided by $L_xL_y$. The close correlation is consistent with radiative losses being supplied primarily by hot-gas enthalpy in both HD and MHD.}
    \label{fig:cooling_vs_Area_vdiff}
\end{figure}

\begin{figure}
    \centering
    \includegraphics[width=\columnwidth]{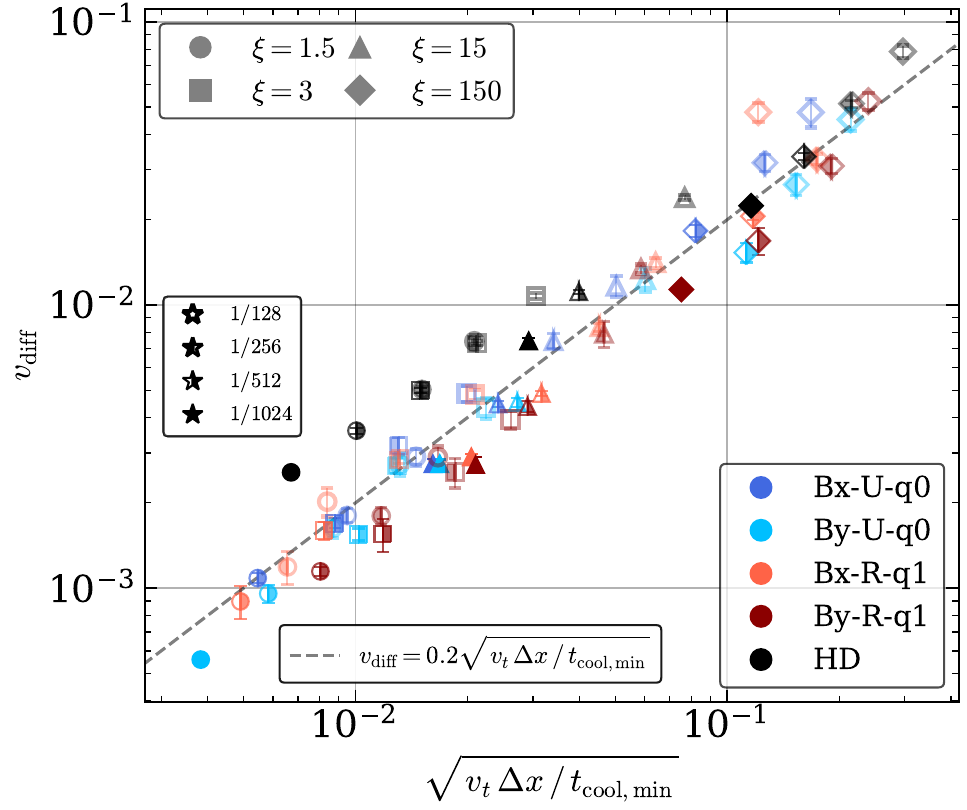}
    \caption{Diffusion velocity measured across the $T=T_{\rm pk}$ surface compared to the value expected from turbulent diffusion on the grid scale. The measured values broadly follow $v_{\rm diff}\sim(v_t\Delta x/t_{\rm cool})^{1/2}$ over a range of resolutions and magnetic configurations.} 
    \label{fig:vdiff_vs_vturb}
\end{figure}

In \Cref{fig:cooling_vs_Area_vdiff}, we compare the net cooling rate to the enthalpy supply estimated from the area of the $T_{\rm pk}$ iso-surface and its grid-scale diffusion velocity. In \Cref{fig:vdiff_vs_vturb}, we compare the diffusion velocity, $v_{\rm diff}$, measured across the same iso-surface on the grid scale (see \S\ref{subsec:time_evol_fid} for a description of our measurement technique), to the expected turbulent diffusion velocity,
\begin{equation}
    v_{\rm diff,exp}\sim \left(\frac{v_t\Delta x}{t_{\rm cool}}\right)^{1/2}.
\end{equation}
For this comparison, we include simulations with resolutions ranging from $\Delta x=1/128$ to $1/1024$.

Similar to \citetalias{Lancaster2026a}, we find a close correspondence between the net cooling rate and the estimated enthalpy supply (\Cref{fig:cooling_vs_Area_vdiff}), consistent with radiative losses being supplied primarily by hot-gas enthalpy. The polarity-reversing runs follow the same relation, providing no evidence here for an additional dominant heating contribution from reconnection.

The measured $v_{\rm diff}$ also follows the expected scaling with $(v_t\Delta x/t_{\rm cool})^{1/2}$ over a range of resolutions, although with some scatter. For a fixed simulation setup, meaning fixed $\xi$ and field configuration, the approximate scaling $v_{\rm diff}\propto \Delta x^{1/2}$ holds reasonably well, consistent with \citetalias{Lancaster2026a}. We suspect that much of the scatter comes from our usage of $v_t$ as the effective turbulent velocity at the interface.  
Nevertheless, the clear correlation, despite this scatter, supports the interpretation that diffusion across the cooling interface is numerical, with a scaling similar to that found by \citetalias{Lancaster2026a}.

\section{Alternative normalizations of the cooling rate}
\label{sec:appendix_alt_norm}
\begin{figure*}
    \centering
    \includegraphics[width=2.0\columnwidth]{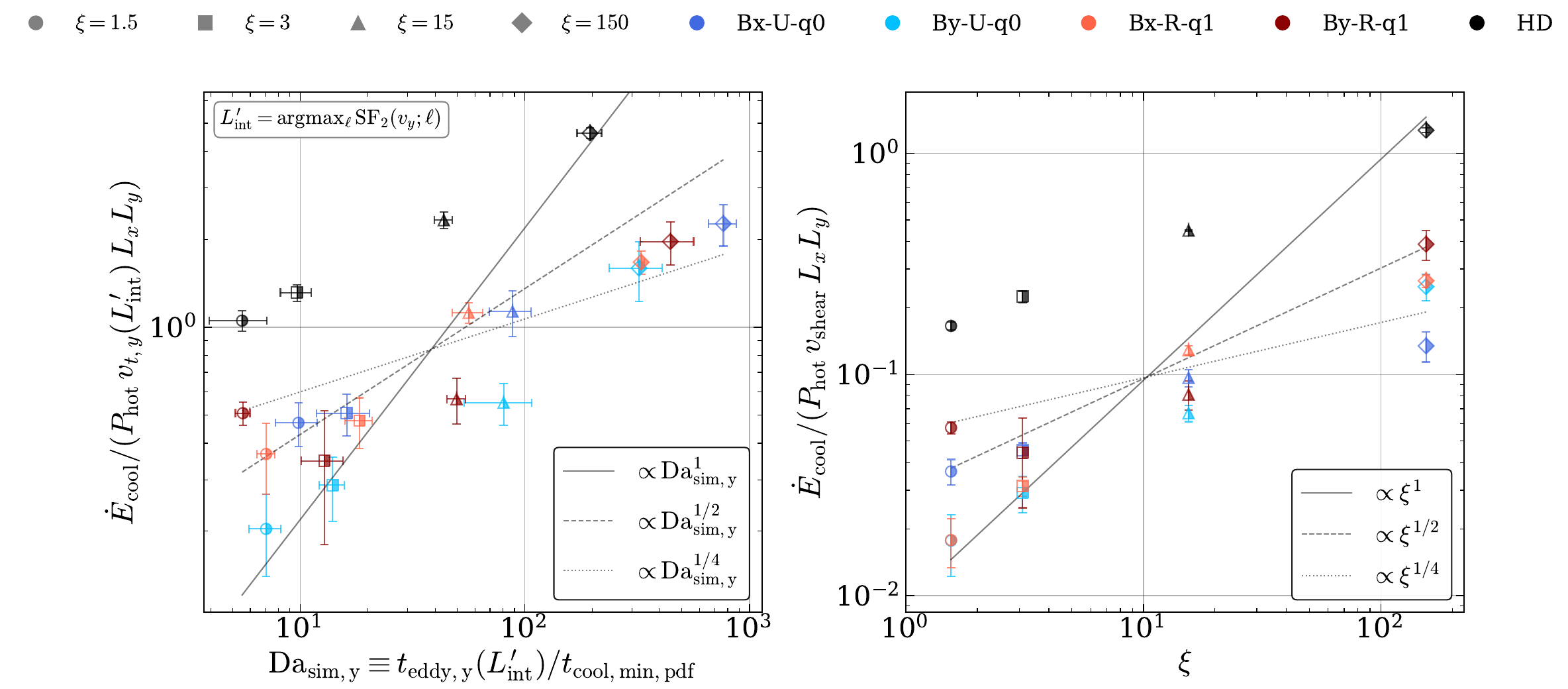}
    \caption{\emph{Left panel:} Same as the upper panel of \Cref{fig:cooling_vs_Da_diff_xi_512}, but using the scaled $y$ component of the velocity field $v_{t,y}$ instead of $v_t$ in the main text, and a different definition of the integral scale, shown in the top left of the figure. This $v_{t,y}$ is also used in calculating $t_{\rm eddy,y}$. The cooling time is estimated from the density--temperature PDF. \emph{Right panel:} Here we use $v_{\rm shear}$ instead, and show this as a function of $\xi=t_{\rm shear}/t_{\rm cool,min}$, which are all input parameters.} 
    \label{fig:cool_Da_diff_norm}
\end{figure*}

The comparison in \S\ref{subsec:diff_xi} uses a measured turbulent velocity, both to normalize the cooling rate and to set the mixing time entering the Damk\"ohler number. Different conventions have been adopted in the literature, and we show two of them here so that our results can be placed alongside earlier work.

The left panel of \Cref{fig:cool_Da_diff_norm} uses the structure function of the raw $y$ velocity, without subtracting its horizontally averaged profile. Following the velocity and scale definitions of \citetalias{Lancaster2026a} and \citetalias{Lancaster2026b}, we define
\begin{align}
    L_{\rm int}'&\equiv
    \operatorname*{argmax}_{\ell}\mathrm{SF}_2(v_y;\ell),\\
    v_{t,y}&\equiv
    \left[3\,\mathrm{SF}_2(v_y;L_{\rm int}')\right]^{1/2},\\
    \mathrm{Da}_{\rm sim,y}&\equiv
    \frac{L_{\rm int}'}{v_{t,y}t_{\rm cool,min,pdf}}.
\end{align}
Here $t_{\rm cool,min,pdf}$ is the minimum positive cooling time obtained by evaluating the cooling prescription over bins of the time-averaged density--temperature PDF. This panel is therefore the version of the comparison that can be read directly against their HD results. 
In the right panel, we normalize the cooling rate by the shear velocity, $v_{\rm shear}$, and plot against $\xi=t_{\rm shear}/t_{\rm cool,min}$. Both are input parameters, fixed by the initial conditions, so this version is comparable with the many earlier TRML studies that report cooling rates in terms of the imposed shear and cooling times rather than a measured turbulent velocity \citep[e.g.][]{Tan2021MNRASa}. 
The suppression of cooling in MHD relative to HD is present in both panels, as it is in \Cref{fig:cooling_vs_Da_diff_xi_512}, so it does not depend on the choice of normalization. 

\section{Evolution of the cooling rate at different resolutions}
\label{sec:appendix_res_history}
\begin{figure*}
    \centering
    \includegraphics[width=2.0\columnwidth]{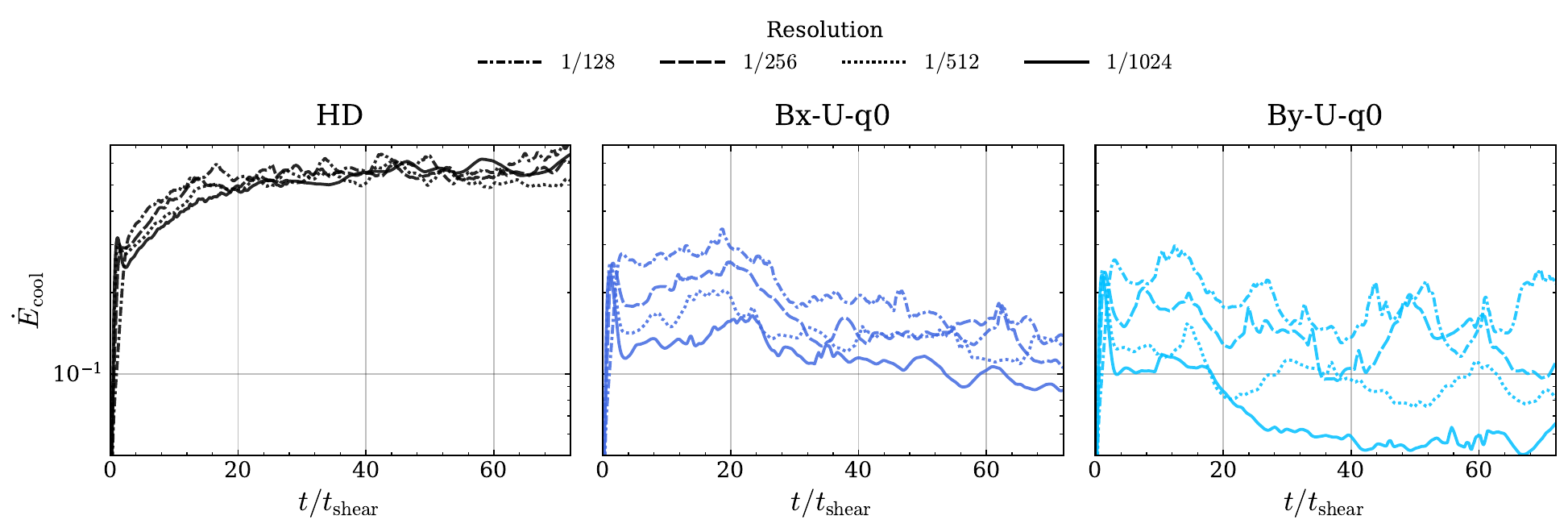}
    \caption{Time evolution of the net cooling rate at four resolutions for the \texttt{HD}, \texttt{Bx-U-q0} and \texttt{By-U-q0} runs.} 
    \label{fig:cool_rate_uni_diff_res}
\end{figure*}

\Cref{fig:cool_rate_uni_diff_res} shows the net cooling rate at each resolution for \texttt{HD}, \texttt{Bx-U-q0} and \texttt{By-U-q0}. The \texttt{HD} curves broadly agree across resolutions. In both MHD configurations, the late-time mean cooling rate decreases with increasing resolution, and at $\Delta x=1/1024$ \texttt{By-U-q0} declines by a further factor of $\sim2$ between $t\approx18$ and $35\,t_{\rm shear}$, more steeply than at lower resolution.


\bsp	
\label{lastpage}
\end{document}